\documentclass[journal]{IEEEtran}

\def\showdisclaimer{}

\RequirePackage{amsmath,amsfonts}
\RequirePackage{algorithmic}
\RequirePackage{algorithm}
\RequirePackage{array}
\RequirePackage{url}
\RequirePackage{verbatim}
\RequirePackage{graphicx}

\RequirePackage{xcolor}
\PassOptionsToPackage{detect-all, per-mode=repeated-symbol, range-units=single, range-phrase=~to~}{siunitx}
\RequirePackage{siunitx}
\PassOptionsToPackage{acronyms,nohypertypes={acronym},nomain}{glossaries}
\RequirePackage{glossaries}
\RequirePackage[pscoord]{eso-pic}
\RequirePackage{multirow}
\RequirePackage{makecell}
\RequirePackage{graphicx}
\RequirePackage{threeparttable}
\RequirePackage{booktabs}
\RequirePackage{colortbl}
\RequirePackage[verbose]{placeins}
\RequirePackage{fancyhdr}
\RequirePackage[hidelinks]{hyperref}
\RequirePackage{MnSymbol}
\RequirePackage{wasysym}
\RequirePackage{xspace}
\RequirePackage{lipsum}
\RequirePackage{soul}
\RequirePackage{calc}
\RequirePackage{orcidlink}
\RequirePackage[noabbrev,capitalise]{cleveref}
\RequirePackage{cite}
\usepackage{adjustbox}
\usepackage{multirow}
\usepackage{ragged2e}
\usepackage{paralist}
\usepackage{float}
\usepackage{placeins}
\usepackage{float}
\usepackage{arydshln}
\usepackage[position=b]{subcaption}
\usepackage{balance}
\usepackage{pdflscape}
\usepackage{afterpage}
\usepackage{circledsteps}

\definecolor{ieee-bright-dblue-100}{rgb}{0.0, 0.3828, 0.6055}
\definecolor{ieee-bright-dblue-80}{rgb}{0.0, 0.4883, 0.6797}
\definecolor{ieee-bright-dblue-60}{rgb}{0.3633, 0.6094, 0.7617}
\definecolor{ieee-bright-dblue-40}{rgb}{0.5898, 0.7383, 0.8398}
\definecolor{ieee-bright-dblue-20}{rgb}{0.8906, 0.8984, 0.9219}
\definecolor{ieee-bright-red-100}{rgb}{0.7266, 0.0469, 0.1836}
\definecolor{ieee-bright-red-80}{rgb}{0.832, 0.3164, 0.3281}
\definecolor{ieee-bright-red-60}{rgb}{0.8906, 0.4922, 0.4805}
\definecolor{ieee-bright-red-40}{rgb}{0.9336, 0.6562, 0.6406}
\definecolor{ieee-bright-red-20}{rgb}{0.9688, 0.8203, 0.8125}
\definecolor{ieee-bright-orange-100}{rgb}{0.9961, 0.6367, 0.0}
\definecolor{ieee-bright-orange-80}{rgb}{0.9844, 0.6953, 0.3125}
\definecolor{ieee-bright-orange-60}{rgb}{0.9883, 0.7695, 0.4844}
\definecolor{ieee-bright-orange-40}{rgb}{0.9922, 0.8359, 0.6562}
\definecolor{ieee-bright-orange-20}{rgb}{0.9961, 0.9219, 0.8164}
\definecolor{ieee-bright-yellow-100}{rgb}{0.9961, 0.8164, 0.0}
\definecolor{ieee-bright-yellow-80}{rgb}{0.9961, 0.8477, 0.2148}
\definecolor{ieee-bright-yellow-60}{rgb}{0.9961, 0.875, 0.4492}
\definecolor{ieee-bright-yellow-40}{rgb}{0.9961, 0.9062, 0.6328}
\definecolor{ieee-bright-yellow-20}{rgb}{0.9961, 0.9531, 0.8125}
\definecolor{ieee-bright-lgreen-100}{rgb}{0.4688, 0.7422, 0.125}
\definecolor{ieee-bright-lgreen-80}{rgb}{0.5742, 0.7852, 0.332}
\definecolor{ieee-bright-lgreen-60}{rgb}{0.6875, 0.8398, 0.5039}
\definecolor{ieee-bright-lgreen-40}{rgb}{0.793, 0.8906, 0.6641}
\definecolor{ieee-bright-lgreen-20}{rgb}{0.8945, 0.9414, 0.8281}
\definecolor{ieee-bright-dgreen-100}{rgb}{0.0, 0.5156, 0.2383}
\definecolor{ieee-bright-dgreen-80}{rgb}{0.1641, 0.6055, 0.3867}
\definecolor{ieee-bright-dgreen-60}{rgb}{0.3906, 0.6953, 0.5234}
\definecolor{ieee-bright-dgreen-40}{rgb}{0.6094, 0.8008, 0.6719}
\definecolor{ieee-bright-dgreen-20}{rgb}{0.8047, 0.8945, 0.8359}
\definecolor{ieee-bright-purple-100}{rgb}{0.5938, 0.1133, 0.5898}
\definecolor{ieee-bright-purple-80}{rgb}{0.6992, 0.3281, 0.668}
\definecolor{ieee-bright-purple-60}{rgb}{0.7812, 0.4961, 0.7461}
\definecolor{ieee-bright-purple-40}{rgb}{0.8555, 0.6602, 0.8281}
\definecolor{ieee-bright-purple-20}{rgb}{0.9219, 0.8281, 0.9023}
\definecolor{ieee-bright-lblue-100}{rgb}{0.0, 0.6094, 0.6484}
\definecolor{ieee-bright-lblue-80}{rgb}{0.0, 0.6797, 0.7188}
\definecolor{ieee-bright-lblue-60}{rgb}{0.2109, 0.75, 0.7812}
\definecolor{ieee-bright-lblue-40}{rgb}{0.5469, 0.8242, 0.8438}
\definecolor{ieee-bright-lblue-20}{rgb}{0.7695, 0.918, 0.9219}
\definecolor{ieee-bright-cyan-100}{rgb}{0.0, 0.707, 0.8828}
\definecolor{ieee-bright-cyan-80}{rgb}{0.0, 0.7227, 0.9453}
\definecolor{ieee-bright-cyan-60}{rgb}{0.2656, 0.7812, 0.957}
\definecolor{ieee-bright-cyan-40}{rgb}{0.5547, 0.8438, 0.9688}
\definecolor{ieee-bright-cyan-20}{rgb}{0.7773, 0.9141, 0.9805}
\definecolor{ieee-bright-white-100}{rgb}{0.9961, 0.9961, 0.9961}
\definecolor{ieee-bright-white-80}{rgb}{0.9961, 0.9961, 0.9961}
\definecolor{ieee-bright-white-60}{rgb}{0.9961, 0.9961, 0.9961}
\definecolor{ieee-bright-white-40}{rgb}{0.9961, 0.9961, 0.9961}
\definecolor{ieee-bright-white-20}{rgb}{0.9961, 0.9961, 0.9961}
\definecolor{ieee-dark-red-100}{rgb}{0.5234, 0.1211, 0.2539}
\definecolor{ieee-dark-red-80}{rgb}{0.6445, 0.2812, 0.3828}
\definecolor{ieee-dark-red-60}{rgb}{0.7422, 0.4727, 0.5234}
\definecolor{ieee-dark-red-40}{rgb}{0.832, 0.6445, 0.6758}
\definecolor{ieee-dark-red-20}{rgb}{0.918, 0.8203, 0.832}
\definecolor{ieee-dark-orange-100}{rgb}{0.9062, 0.4648, 0.1328}
\definecolor{ieee-dark-orange-80}{rgb}{0.9648, 0.5664, 0.3164}
\definecolor{ieee-dark-orange-60}{rgb}{0.9766, 0.6758, 0.4805}
\definecolor{ieee-dark-orange-40}{rgb}{0.9844, 0.7773, 0.6523}
\definecolor{ieee-dark-orange-20}{rgb}{0.9922, 0.8789, 0.8125}
\definecolor{ieee-dark-yellow-100}{rgb}{0.9961, 0.7773, 0.1719}
\definecolor{ieee-dark-yellow-80}{rgb}{0.9961, 0.8086, 0.375}
\definecolor{ieee-dark-yellow-60}{rgb}{0.9961, 0.875, 0.4492}
\definecolor{ieee-dark-yellow-40}{rgb}{0.9961, 0.8984, 0.6875}
\definecolor{ieee-dark-yellow-20}{rgb}{0.9961, 0.9453, 0.8438}
\definecolor{ieee-dark-lgreen-100}{rgb}{0.3945, 0.5508, 0.0938}
\definecolor{ieee-dark-lgreen-80}{rgb}{0.5078, 0.6289, 0.293}
\definecolor{ieee-dark-lgreen-60}{rgb}{0.6367, 0.7188, 0.4688}
\definecolor{ieee-dark-lgreen-40}{rgb}{0.7539, 0.8047, 0.6367}
\definecolor{ieee-dark-lgreen-20}{rgb}{0.875, 0.9023, 0.8125}
\definecolor{ieee-dark-dgreen-100}{rgb}{0.0, 0.3867, 0.2539}
\definecolor{ieee-dark-dgreen-80}{rgb}{0.1836, 0.5, 0.3906}
\definecolor{ieee-dark-dgreen-60}{rgb}{0.3984, 0.6172, 0.5273}
\definecolor{ieee-dark-dgreen-40}{rgb}{0.5938, 0.7422, 0.6758}
\definecolor{ieee-dark-dgreen-20}{rgb}{0.793, 0.8711, 0.8359}
\definecolor{ieee-dark-purple-100}{rgb}{0.4648, 0.1445, 0.5117}
\definecolor{ieee-dark-purple-80}{rgb}{0.5898, 0.3242, 0.6016}
\definecolor{ieee-dark-purple-60}{rgb}{0.6914, 0.4883, 0.6953}
\definecolor{ieee-dark-purple-40}{rgb}{0.7969, 0.6523, 0.793}
\definecolor{ieee-dark-purple-20}{rgb}{0.8945, 0.8203, 0.8945}
\definecolor{ieee-dark-cyan-100}{rgb}{0.0, 0.4492, 0.4648}
\definecolor{ieee-dark-cyan-80}{rgb}{0.0, 0.5469, 0.5664}
\definecolor{ieee-dark-cyan-60}{rgb}{0.3047, 0.6602, 0.668}
\definecolor{ieee-dark-cyan-40}{rgb}{0.5586, 0.7695, 0.7734}
\definecolor{ieee-dark-cyan-20}{rgb}{0.7734, 0.8789, 0.8789}
\definecolor{ieee-dark-dblue-100}{rgb}{0.0, 0.1562, 0.332}
\definecolor{ieee-dark-dblue-80}{rgb}{0.1797, 0.3008, 0.4609}
\definecolor{ieee-dark-dblue-60}{rgb}{0.3828, 0.4609, 0.5859}
\definecolor{ieee-dark-dblue-40}{rgb}{0.5781, 0.6289, 0.7188}
\definecolor{ieee-dark-dblue-20}{rgb}{0.7852, 0.8047, 0.8555}
\definecolor{ieee-dark-grey-100}{rgb}{0.457, 0.4688, 0.4805}
\definecolor{ieee-dark-grey-80}{rgb}{0.5625, 0.5625, 0.5742}
\definecolor{ieee-dark-grey-60}{rgb}{0.6641, 0.6641, 0.6758}
\definecolor{ieee-dark-grey-40}{rgb}{0.7734, 0.7695, 0.7773}
\definecolor{ieee-dark-grey-20}{rgb}{0.8789, 0.8828, 0.8828}
\definecolor{ieee-dark-black-100}{rgb}{0.0, 0.0, 0.0}
\definecolor{ieee-dark-black-80}{rgb}{0.3438, 0.3477, 0.3555}
\definecolor{ieee-dark-black-60}{rgb}{0.5, 0.5078, 0.5195}
\definecolor{ieee-dark-black-40}{rgb}{0.6523, 0.6602, 0.6719}
\definecolor{ieee-dark-black-20}{rgb}{0.8164, 0.8242, 0.8281}

\newacronym{ai}{AI}{Artificial Intelligence}
\newacronym{asic}{ASIC}{application-specific integrated circuit}
\newacronym{axi4}{AXI4}{advanced eXtensible interface 4}
\newacronym{bmc}{BMC}{baseboard management controller}
\newacronym{cpu}{CPU}{central processing unit}
\newacronym{dm}{DM}{data mover}
\newacronym{dma}{DMA}{direct memory access}
\newacronym{dmac}{DMAC}{direct memory access controller}
\newacronym{dmae}{DMAE}{direct memory access engine}
\newacronym{dram}{DRAM}{dynamic random access memory}
\newacronym{dsp}{DSP}{digital signal processing}
\newacronym{dvfs}{DVFS}{dynamic voltage and frequency scaling}
\newacronym{e2e}{E2E}{end-to-end}
\newacronym{fifo}{FIFO}{first in, first out}
\newacronym{fpga}{FPGA}{field programmable gate array}
\newacronym{fpu}{FPU}{floating-point unit}
\newacronym{frep}{FREP}{Floating-point Repetition}
\newacronym{gemm}{GEMM}{general matrix multiply}
\newacronym{gp}{GP}{general-purpose}
\newacronym{hbm}{HBM}{high-bandwidth memory}
\newacronym{hil}{HIL}{hardware-in-the-loop}
\newacronym{hlc}{HLC}{high-level controller}
\newacronym{hpc}{HPC}{high-performance computing}
\newacronym{idma}{iDMA}{intelligent DMA}
\newacronym{ihls}{iHLS}{IP-based high-level synthesis}
\newacronym{ip}{IP}{intellectual property}
\newacronym{isa}{ISA}{instruction set architecture}
\newacronym{l1}{L1}{level-one}
\newacronym{l2}{L2}{level-two}
\newacronym{l3}{L3}{level-three}
\newacronym{llc}{LLC}{low-level controller}
\newacronym{mcu}{MCU}{microcontroller unit}
\newacronym{mimo}{MIMO}{multi-input multi-output}
\newacronym{ml}{ML}{machine learning}
\newacronym{noc}{NoC}{network-on-chip}
\newacronym{ooc}{OOC}{out-of-context}
\newacronym{obi}{OBI}{open bus protocol}
\newacronym{os}{OS}{operating system}
\newacronym{pcf}{PCF}{power control firmware}
\newacronym{pcs}{PCS}{power controller system}
\newacronym{pdk}{PDK}{process design kit}
\newacronym{pe}{PE}{processing element}
\newacronym{pfct}{PFCT}{periodic frequency control task}
\newacronym{pulp}{PULP}{parallel ultra-low power}
\newacronym{pvct}{PVCT}{periodic voltage control task}
\newacronym{pvt}{PVT}{process-voltage-temperature}
\newacronym{rpc}{RPC}{reduced pin count}
\newacronym{rtl}{RTL}{register transfer level}
\newacronym{sdma}{sDMAE}{sensor DMAE}
\newacronym{soa}{SoA}{state-of-the-art}
\newacronym[longplural=systems on chip]{soc}{SoC}{system on chip}
\newacronym{simd}{SIMD}{single instruction, multiple data}
\newacronym[longplural=scratchpad memories]{spm}{SPM}{scratchpad memory}
\newacronym{spmv}{SpMV}{sparse matrix-vector multiply}
\newacronym{spmm}{SpMM}{sparse matrix-matrix multiply}
\newacronym{spmspm}{SpMSpM}{sparse-sparse matrix multiply}
\newacronym{sram}{SRAM}{static read-only memory}
\newacronym{ssr}{SSR}{Stream Semantic Register}
\newacronym{tcdm}{TCDM}{tightly-coupled data memory}
\newacronym{ulp}{ULP}{ultra-low-power}
\newacronym{vrm}{VRM}{voltage regulator module}
\newacronym{mpc}{MPC}{model predictive control}
\newacronym{nn}{NN}{neural network}
\newacronym{la}{LA}{linear algebra}
\newacronym{llm}{LLM}{large language model}
\newacronym{su}{SU}{streaming unit}
\newacronym{d2d}{D2D}{die-to-die}
\newacronym{c2c}{C2C}{chip-to-chip}
\newacronym{grli}{GRLI}{globally regular, locally irregular}
\newacronym{gpt}{GPT}{generative pretrained transformer}
\newacronym{gcn}{GCN}{graph convolutional network}
\newacronym{gsm}{GSM}{Graduate Student Member}
\newacronym{fma}{FMA}{fused multiply-accumulate}
\newacronym{iotlb}{IOTLB}{IO translation lookaside buffer}
\newacronym{phy}{PHY}{physical layer}
\newacronym{ddr}{DDR}{double-data-rate}
\newacronym{zif}{ZIF}{zero insertion force}
\newacronym{ate}{ATE}{automatic test equipment}
\newacronym{fll}{FLL}{frequency-locked loop}
\newacronym{flop}{FLOP}{floating-point operation}
\newacronym{dsa}{DSA}{domain-specific accelerator}
\newacronym{cgra}{CGRA}{coarse-grained reconfigurable array}

\newcommand{\glsf}[1]{\glsreset{#1}\gls{#1}}

\newcommand{\etal}{\emph{et al.}}
\newcommand{\x}{$\times$}

\newcommand{\gsm}{{\gls{gsm}}}
\newcommand*\sqmm{\si{\milli\metre\squared}}

\DeclareSIUnit{\x}{\!\ensuremath{\times}}
\DeclareSIUnit\bit{b}
\DeclareSIUnit\bps{bps}
\DeclareSIUnit\flop{FLOP}
\DeclareSIUnit\dash{\text{-}}
\DeclareSIUnit\gateeq{GE}
\DeclareSIUnit\comp{COMP}
\DeclareSIUnit\tok{tok}
\newcommand{\gf}{{GlobalFoundries}}
\newcommand{\gfs}{{GlobalFoundries'}}

\newcommand{\occamy}{{Occamy}}
\newcommand{\riscv}{\mbox{RISC-V}}

\AtBeginDocument{\DeclareCaptionSubType{figure}}
\crefname{subfigure}{figure}{figures}
\Crefname{subfigure}{Figure}{Figures}

\ifx\showrebuttal\undefined
    \newcommand\blackcircle[1]{\Circled[inner color=white, outer color=white, fill color=black]{#1}}
\else
    \newcommand\blackcircle[1]{\Circled[inner color=white, outer color=white, fill color=blue]{#1}}
\fi

\def\reviewpass{v2}

\def\thetitle{Occamy: A 432-Core Dual-Chiplet Dual-HBM2E 768-DP-GFLOP/s RISC-V System for 8-to-64-bit Dense and Sparse Computing in 12nm FinFET}

\begin{document}

\ifx\showdisclaimer\undefined
\else
\AddToShipoutPictureBG*{%
  \AtPageUpperLeft{%
    \hspace{\paperwidth}%
    \raisebox{-\baselineskip}{%
      \makebox[-35pt][r]{\footnotesize{
        \copyright~2025~IEEE. Personal use of this material is permitted. %
        Permission from IEEE must be obtained for all other uses, in any current or future media, including
      }}
}}}%

\AddToShipoutPictureBG*{%
  \AtPageUpperLeft{%
    \hspace{\paperwidth}%
    \raisebox{-2\baselineskip}{%
      \makebox[-37pt][r]{\footnotesize{
        reprinting/republishing this material for advertising or promotional purposes, creating new collective works, for resale or
      }}
}}}%

\AddToShipoutPictureBG*{%
  \AtPageUpperLeft{%
    \hspace{\paperwidth}%
    \raisebox{-3\baselineskip}{%
      \makebox[-185pt][r]{\footnotesize{
       reuse of any copyrighted component of this work in other works.
      }}
}}}%
\fi

\title{\thetitle}
\ifx\showrevision\undefined
    \newcommand{\todo}[1]{{#1}}
\else
    \newcommand{\todo}[1]{{\textcolor{red}{#1}}}
    \AddToShipoutPictureFG{%
        \put(%
            8mm,%
            \paperheight-1.5cm%
            ){\vtop{{\null}\makebox[0pt][c]{%
                \rotatebox[origin=c]{90}{%
                    \huge\textcolor{red!75}{\reviewpass}%
                }%
            }}%
        }%
    }
    \AddToShipoutPictureFG{%
        \put(%
            \paperwidth-6mm,%
            \paperheight-1.5cm%
            ){\vtop{{\null}\makebox[0pt][c]{%
                \rotatebox[origin=c]{90}{%
                    \huge\textcolor{red!30}{ETH Zurich - Unpublished - Confidential - Draft - Copyright Paul, Thomas 2024}%
                }%
            }}%
        }%
    }
\fi

\ifx\showrebuttal\undefined
    \newcommand{\rev}[1]{#1}
\else
    \newcommand{\rev}[1]{{\textcolor{blue}{#1}}}
\fi

\ifx\showrebuttal\undefined
    \newenvironment{revenv}{}{}
\else
    \newenvironment{revenv}{\color{blue}}{\color{black}}
\fi

\ifx\showrebuttal\undefined
    \newcommand{\revdel}[1]{}
\else
    \newcommand{\revdel}[1]{\textcolor{ieee-bright-red-100}{\st{#1}}}
\fi

\ifx\showrebuttal\undefined
    \newcommand{\revrep}[2]{#2}
\else
    \newcommand{\revrep}[2]{\revdel{#1} \rev{#2}}
\fi

\ifx\showrebuttal\undefined
    \newcommand{\revprg}[1]{}
\else
    \newcommand{\revprg}[1]{\hspace{-0.5ex}\textcolor{red}{\scalebox{.2}[1.5]{$\blacksquare$}}\hspace{-0.5ex}}
\fi

\ifx\showchecknums\undefined
    \newcommand{\cn}[1]{#1}
\else
    \newcommand{\cn}[1]{\textcolor{blue}{#1}}
\fi

\author{
    Paul~Scheffler\orcidlink{0000-0003-4230-1381}\IEEEauthorrefmark{1},~\IEEEmembership{\glsf{gsm}, IEEE},
    Thomas~Benz\orcidlink{0000-0002-0326-9676}\IEEEauthorrefmark{1},~\IEEEmembership{\gsm{}, IEEE},
    Viviane~Potocnik\orcidlink{0009-0004-9412-6081},~\IEEEmembership{\gsm{}, IEEE},
    Tim~Fischer\orcidlink{0009-0007-9700-1286},~\IEEEmembership{\gsm{}, IEEE},
    Luca~Colagrande\orcidlink{0000-0002-7986-1975},~\IEEEmembership{\gsm{}, IEEE},
    Nils~Wistoff\orcidlink{0000-0002-8683-8060},~\IEEEmembership{\gsm{}, IEEE},
    Yichao~Zhang\orcidlink{0009-0008-7508-599X},~\IEEEmembership{\gsm{}, IEEE},
    Luca~Bertaccini\orcidlink{0000-0002-3011-6368},~\IEEEmembership{\gsm{}, IEEE},
    Gianmarco~Ottavi\orcidlink{0000-0003-0041-7917},~\IEEEmembership{\gsm{}, IEEE},
    Manuel~Eggimann\orcidlink{0000-0001-8395-7585},~\IEEEmembership{Member, IEEE},
    Matheus~Cavalcante\orcidlink{0000-0001-9199-1708},~\IEEEmembership{Member, IEEE},
    Gianna~Paulin\orcidlink{0000-0002-1310-0911},~\IEEEmembership{Member, IEEE},
    Frank~K.~G\"{u}rkaynak\orcidlink{0000-0002-8476-554X},
    Davide~Rossi\orcidlink{0000-0002-0651-5393},~\IEEEmembership{Senior Member, IEEE},
    and~Luca~Benini\orcidlink{0000-0001-8068-3806},~\IEEEmembership{Fellow,~IEEE}%
    \IEEEcompsocitemizethanks{%
    \noindent%
    \IEEEauthorrefmark{1} Authors contributed equally to this research.\protect\\
    \IEEEcompsocthanksitem P.~Scheffler, T.~Benz, V.~Potocnik, T.~Fischer, L.~Colagrande, N.~Wistoff, Y.~Zhang,  L.~Bertaccini, F.~K.~G\"{u}rkaynak, and L.~Benini are with the Integrated Systems Laboratory (IIS), ETH Zurich, Switzerland (email: paulsc@ethz.ch).\protect\\
    G.~Ottavi, D.~Rossi, and L.~Benini are with the Department of Electrical, Electronic and Information Engineering (DEI), University of Bologna, Italy.\protect\\
    M.~Eggimann and G.~Paulin are with Axelera AI.\protect\\
    M.~Cavalcante is with the Robust Systems Group (RSG) at the Computer Systems Laboratory (CSL) of Stanford University, CA, USA.\protect\\%
    }%
}

\ifx\showdisclaimer\undefined
\markboth{IEEE Journal on Solid-State Circuits (JSSC)}%
{Scheffler \MakeLowercase{\etal}: \thetitlesingleline}
\else
\fi

\maketitle

\begin{abstract}
ML and HPC applications increasingly combine dense and sparse memory access computations to maximize storage efficiency. 
However, existing CPUs and GPUs struggle to flexibly handle these heterogeneous workloads with consistently high compute efficiency.
We present Occamy, a 432-Core, 768-DP-GFLOP/s, dual-HBM2E, dual-chiplet RISC-V system with a latency-tolerant hierarchical interconnect and in-core streaming units (SUs) designed to accelerate dense and sparse FP8-to-FP64 ML and HPC workloads.
We implement Occamy's compute chiplets in \SI{12}{\nano\meter} FinFET, and its passive interposer, Hedwig, in a \SI{65}{\nano\meter} node.
On dense linear algebra (LA), Occamy achieves a competitive FPU utilization of \cn{89\%}. %
On stencil codes, Occamy reaches an FPU utilization of \cn{83\%} and a technology-node-normalized compute density of \cn{\SI{11.1}{DP\text{-}\giga\flop\per\second\per\milli\meter^2}}, leading \gls{soa} processors by \cn{1.7\x}~and \rev{\cn{1.2\x}}, respectively.
On sparse-dense \gls{la}, it achieves \cn{42\%} FPU utilization and a normalized compute density of \cn{\SI{5.95}{DP\text{-}\giga\flop\per\second\per\milli\meter^2}}, surpassing the \gls{soa} by \cn{5.2\x}~and \cn{11\x}, respectively.
\rev{On, sparse-sparse LA, Occamy reaches} a throughput of up to \cn{\SI{187}{\giga\comp\per\second}} at \cn{\SI{17.4}{\giga\comp\per\second\per\watt}} and a compute density of \cn{\SI{3.63}{\giga\comp\per\second\per\milli\meter^2}}.
\rev{Finally, we reach up to \SI{75}{\percent} and \SI{54}{\percent} FPU utilization on and dense (LLM) and graph-sparse (GCN) ML inference workloads.} 
Occamy's RTL is freely available under a permissive open-source license.

\end{abstract}


\glsresetall

\section{Introduction}
\label{sec:intro}

Surging performance needs and large, sparse models in \gls{ml} and \gls{hpc} push data-driven workloads toward a mixture of sparse and dense computations.
In \gls{ml}, the sparsification of dense models significantly reduces both their computational and memory footprints~\cite{hoefler2021sparsity}, and the recent proliferation of \textgreater 100B-parameter transformers has reinvigorated sparsification efforts to address severe memory capacity bottlenecks~\cite{dettmers2023spqr, xia2023flash}.
In \gls{hpc}, applications like multiphysics simulation and graph analytics heavily rely on sparse \gls{la}, stencil computations, and graph pattern matching to handle large, high-dimensional problems~\cite{Zhang2025Survey}.

Large-scale \gls{ml} and \gls{hpc} applications traditionally target manycore CPUs and GPUs whose parallel datapaths achieve high ($\geq$\SI{75}{\percent}) FPU utilization and energy efficiency on dense workloads~\cite{nvidia_a100_dense, siegmann2024first}.
However, these general-purpose architectures struggle with the indirect address computations, irregular memory access patterns, and computational imbalance inherent to sparse workloads.
Consequently, they achieve peak FPU utilizations of less than \cn{\SI{50}{\percent}} on stencil codes~\cite{cfd_amd_epyc_rome_tpds21, hirokawa2022large, nvidia_a100_stencil} and less than \cn{\SI{10}{\percent}} on sparse \gls{la}~\cite{amd_sparse, alappat2020a64fx, nvidia_a100_sparse, siegmann2024first}.
While hardware accelerators for sparse computation have been proposed, they usually target narrow, domain-specific sparsity regimes:
accelerators for \gls{ml}~\cite{huang2025hybr, yue2024cimblock, feng2024pntcloud, zhang2021snap} support low-precision data formats ($\leq$\SI{16}{\bit}) and low ($\leq$\SI{80}{\percent}) or structured sparsity,
while more generic sparse \gls{la}~\cite{Srivastava2020MatRaptorAS, Pal2018OuterSPACEAO, park2020output, sadi2019efficient} and tensor algebra~\cite{Hegde2019ExTensorAA} accelerators often target specific operators~\cite{Srivastava2020MatRaptorAS, Pal2018OuterSPACEAO, park2020output, sadi2019efficient} or dataflows~\cite{Srivastava2020MatRaptorAS, Pal2018OuterSPACEAO}. 
General-purpose acceleration solutions based on \glspl{cgra}~\cite{Dadu2019TowardsGP,koul2024onyx} or \gls{isa} extensions~\cite{siracusa2023tmu, Domingos2021UnlimitedVE, zhu2019sparsetensorcore, Wang2019StreambasedMA, Wang2021DualsideST, Rao2022SparseCoreSI} are more flexible, but they have not yet been demonstrated on silicon at scale~\cite{Dadu2019TowardsGP, siracusa2023tmu, Domingos2021UnlimitedVE, zhu2019sparsetensorcore, Wang2019StreambasedMA, Wang2021DualsideST, Rao2022SparseCoreSI} or tackle only one-~\cite{siracusa2023tmu, Domingos2021UnlimitedVE, zhu2019sparsetensorcore, Wang2019StreambasedMA} or two-sided sparsity~\cite{Wang2021DualsideST, Rao2022SparseCoreSI}.
Hence, there is a lack of \emph{silicon-proven} processors that efficiently handle the diverse sparse compute requirements of current \gls{ml} and \gls{hpc} without compromising on dense compute performance.

\begin{figure*}[ht!]
    \begin{subcaptionblock}{0.318\linewidth}%
        \centering%
        \includegraphics[width=\linewidth]{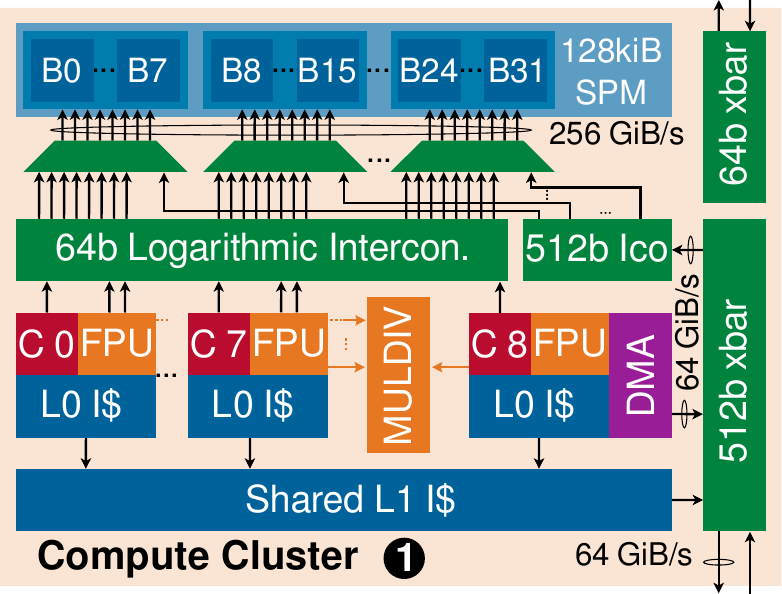}%
        \vspace{-0.4em}%
        \caption{Compute cluster}%
        \vspace{0.25em}%
        \label{fig:arch_cluster}%
        \vfill%
        \centering%
        \includegraphics[width=\linewidth]{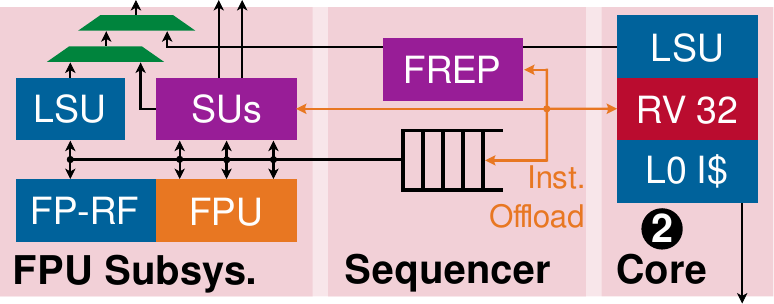}%
        \vspace{-0.3em}%
        \caption{Worker core}%
        \label{fig:arch_cc}%
    \end{subcaptionblock}\hfill
    \begin{subcaptionblock}{0.1836\linewidth}
        \centering%
        \includegraphics[width=\linewidth]{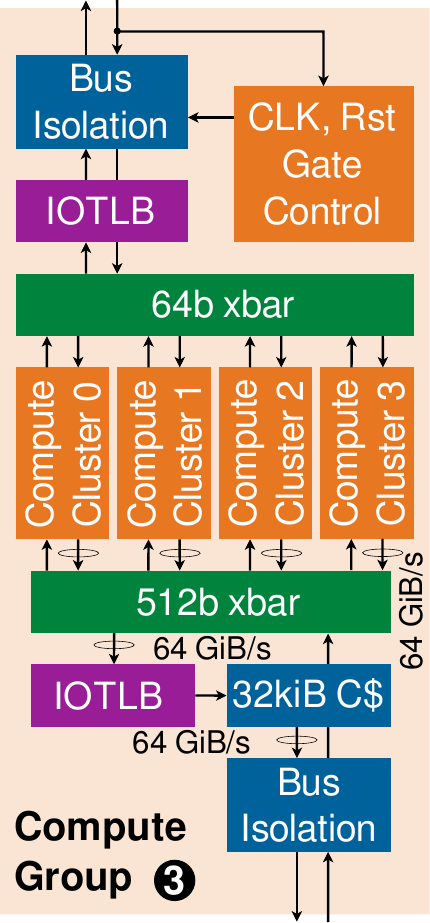}%
        \vspace{-0.3em}%
        \caption{Compute group}%
        \label{fig:arch_group}%
    \end{subcaptionblock}\hfill
    \begin{subcaptionblock}{0.477\linewidth}
        \centering%
        \includegraphics[width=\linewidth]{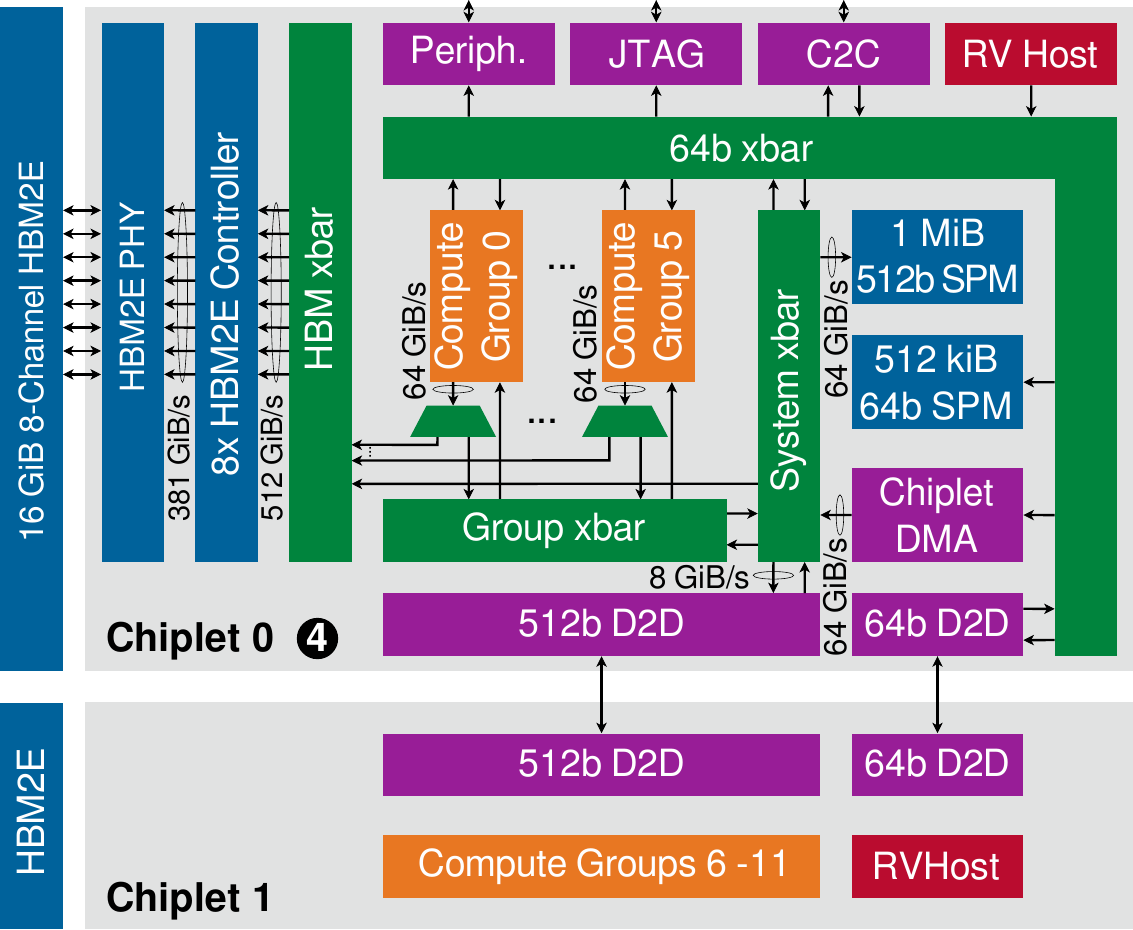}%
        \vspace{-0.2em}%
        \caption{Occamy chiplets}%
        \label{fig:arch_chiplet}%
    \end{subcaptionblock}
    \caption{Hierarchical architecture of the two Occamy compute chiplets including the two HBM2E interfaces and D2D link.} 
    \label{fig:arch_overview}
\end{figure*}

We present \emph{{\occamy}}, a 432-core, 768-\si{DP\dash\giga\flop\per\second} dual-chiplet {\riscv} system whose architecture is specifically designed to flexibly handle both dense and sparse \gls{ml} and \gls{hpc} workloads with high FPU utilization and energy efficiency. 
\rev{Unlike existing hardware targeting sparse compute, Occamy is designed for \emph{general-purpose} sparse acceleration and avoids specialization to specific workloads, sparsity patterns, and value densities. }
Each Occamy chiplet features 24 compute \emph{clusters}, a Linux-capable \revprg{\riscv}~host, \SI{16}{\gibi\byte} of HBM2E DRAM, and a fully digital \revprg{fault-tolerant \cn{\SI{1.6}{\pico\joule\per\bit}}} \gls{d2d} link.
In each cluster, nine cores share \rev{an} explicitly managed \SI{128}{\kibi\byte} \gls{spm};
eight \emph{worker} cores with SIMD-capable \revprg{FP8-to-FP64} \glsunset{fpu}\glspl{fpu}~\cite{bertaccini2024fpu} are kept busy by~sparsity-capable \glspl{su}~\cite{scheffler2023sssr}, while an additional \emph{DMA control} core~\cite{benz2024idma} enables \revprg{asynchronous,} latency-tolerant \revprg{$\leq$2D} transfers of large data tiles to or from the \gls{spm}.
The hierarchical chiplet interconnect provides a 512-bit network for \rev{bulk} data transfers and a 64-bit network for message passing and synchronization.

We implemented Occamy's \cn{\SI{73}{\milli\meter\squared}} \rev{compute} chiplets in \gfs~\SI{12}{\nano\meter} LP+ \revprg{FinFET} technology and \rev{its} passive interposer named \emph{Hedwig} in a \SI{65}{\nano\meter} node.
On dense workloads, {\occamy} achieves competitive FPU utilizations of up to \cn{\SI{89}{\percent}}. %
Executing stencil codes, we reach FPU utilizations of up to \cn{\SI{83}{\percent}} and a technology-node-normalized compute density of up to \cn{\SI{11.1}{DP\text{-}\giga\flop\per\second\per\milli\meter^2}}, leading \rev{state-of-the-art processors} by \cn{1.7\x}~and 
\rev{\cn{1.2\x}}, respectively. \revprg{, with respect to \gls{soa} CPUs and GPUs.}
On sparse-dense \gls{la}, we achieve up to \cn{\SI{42}{\percent}} FPU utilization and a 
\revprg{technology-} node-normalized compute density of up to \cn{\SI{5.95}{DP\text{-}\giga\flop\per\second\per\milli\meter^2}}, substantially surpassing the \gls{soa} by \cn{5.2\x}~and \cn{11\x}, respectively.
\rev{We further} achieve a sparse-sparse \gls{la} throughput of up to \cn{\SI{187}{\giga\comp\per\second}} at an energy efficiency of \cn{\SI{17.4}{\giga\comp\per\second\per\watt}}  and a compute density of \cn{\SI{3.63}{\giga\comp\per\second\per\milli\meter^2}}.
\rev{Finally, we reach FPU utilizations of up to \SI{75}{\percent} and \SI{54}{\percent} on \glsunset{gpt}\gls{gpt}-J inference and a \gls{gcn} layer, respectively.} 
\revprg{Furthermore, }~To the best of our knowledge, Occamy is the first open-source multi-chiplet RISC-V manycore demonstrated in silicon;
the \gls{rtl} description of its digital core, including the compute groups, hierarchical interconnect, and \gls{d2d} link, is freely available under a permissive open-source license\footnote{\url{https://github.com/pulp-platform/occamy}}.

This paper extends our previous work~\cite{occamy_vlsi_2024}, significantly elaborating on Occamy's architecture, its programming model, and its silicon implementation. We further add new measurement results on dense \gls{la} workloads, FP8-to-FP64 SIMD computation, \rev{\gls{ml} inference applications,} and our \gls{d2d} link and extend our state-of-the-art comparison with additional related works and metrics.

The remainder of the paper is organized as follows. \Cref{sec:arch} presents Occamy's hierarchical compute chiplet architecture. In \Cref{sec:progmodel}, we illustrate how Occamy,  its \glspl{su}, and its asynchronous DMA engines are programmed to achieve high compute efficiency. \Cref{sec:impl} describes how the Occamy chiplets, the Hedwig interposer, and the carrier board were implemented. \Cref{sec:results} presents our experimental results. In \Cref{sec:soacomp},  we discuss how Occamy compares to \gls{soa} CPUs and GPUs as well as existing sparse acceleration proposals. Finally, we provide a conclusion in \Cref{sec:conlusion}.

\section{Architecture}
\label{sec:arch}

Occamy combines two 216-core compute chiplets on a passive interposer, each paired with an eight-device, \SI{16}{\gibi\byte} HBM2E stack and connected through a fully digital, fault-tolerant \gls{d2d} link.
\cref{fig:arch_overview} shows the hierarchical compute chiplet architecture%
, which we describe in a bottom-up fashion in the following sections.

\subsection{Compute Cluster}
\label{sec:arch_cluster}

Occamy's compute cores are based on the RV32G \gls{isa} and are organized into compute \emph{clusters} \blackcircle{1}~\cite{zaruba2020snitch} shown in \Cref{fig:arch_cluster}. 
Within each cluster, eight worker cores and one \glsunset{dma}\gls{dma} control core share a \SI{128}{\kibi\byte} 32-bank \gls{spm} through a single-cycle logarithmic interconnect with double-word interleaving.
The clusters also feature \SI{8}{\kibi\byte} of shared L1 instruction cache, a shared integer multiply-divide unit, a local hardware synchronization barrier, and 16 retargetable performance counters  
capable of tracking various per-core and cluster-wide events.

Each worker core \blackcircle{2}, shown in \Cref{fig:arch_cc}, \rev{features} a 64-bit-wide \glsunset{simd}\gls{simd} \gls{fpu} supporting FP64, FP32, FP16, FP16alt (8,7), FP8, and FP8alt (4,3) formats.
In addition to \gls{fma} instructions, the FPU supports widening sum-dot-product and three-addend summation instructions for all FP8 and FP16 formats~\cite{bertaccini2024fpu}.
Two worker-core ISA extensions maximize the FPU utilization for both regular and irregular workloads: a hardware loop buffer~\cite{zaruba2020snitch} and three sparsity-capable \glspl{su}~\cite{scheffler2023sssr}.

\begin{figure}[t]
\centering
\includegraphics[width=0.95\linewidth]{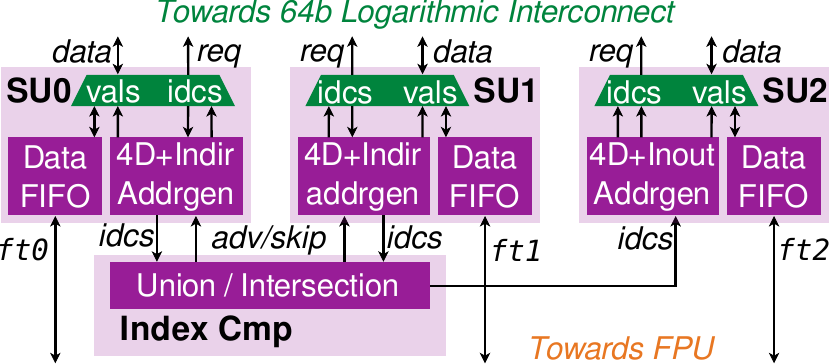}
\caption{Architecture and interconnection of the three cooperating sparsity-capable SUs in each worker core.}
\label{fig:arch_su}
\end{figure}

The \glspl{su}, shown in \cref{fig:arch_su}, map buffered streams of \gls{spm} accesses directly to floating-point registers, generating the necessary addresses in hardware.
All three \glspl{su} in each core support up to 4D strided accesses to accelerate dense tensor workloads\rev{; for instance, the \gls{gemm} used in our later evaluation in \cref{sec:res_perf} utilizes all four \gls{su} loop levels to cover the algorithm's three nested loops and an inner unroll for performance.}
Two \glspl{su} additionally support indirect streams with \mbox{8-,} 16-, or 32-bit indices to accelerate the scatter-gather accesses underlying irregular-access applications such as sparse-dense \gls{la} and stencil codes. 
Finally, these indirect \glspl{su} can also compare their indices to accelerate the sparse tensor intersection and union underlying sparse-sparse \gls{la} or graph matching; the third \gls{su} can optionally write out the joint indices for sparse result tensors.
\rev{
Our \glspl{su} can handle any sparse tensor format whose major axis is given by a value-index array pair, which includes the widespread and scalable CSR, CSC, CSF, and many of their variations, 
without restrictive assumptions on operand structure or density.
}
In general, \rev{our} \glspl{su} enable a sustained per-core bandwidth of up to three double-words per cycle into the shared \gls{spm}.
We will discuss in detail how \glspl{su} are programmed and leveraged for dense and sparse workload acceleration in~\cref{sec:progmodel_wsu}.

\rev{The cluster \gls{spm} is dimensioned to balance throughput, interconnect complexity, and area.}
\rev{32 banks are chosen to reduce the probability of banking conflicts between \glspl{su} (24 per cluster) while keeping the logarithmic interconnect physically implementable.}
\rev{A capacity of \SI{128}{\kibi\byte} achieves a reasonable \glsunset{sram}\gls{sram} bit density (avoiding significant drops that would appear for $\leq$\SI{4}{\kibi\byte} banks) while keeping area prevalently allocated to compute logic.}

The \gls{dma} control core in each cluster features a tightly-coupled 512-bit \gls{dma} engine~\cite{benz2024idma}, enabling asynchronous $\leq$2D transfers between external memory (other clusters, HBM2E, or other chiplet) and the local \gls{spm}.
This core coordinates the computation of worker cores and their fine-grained, low-latency accesses to the \gls{spm} with the latency-tolerant \gls{dma} transfers of large, double-buffered data tiles to and from the \gls{spm}.
\rev{The \gls{dma} engine accesses} blocks of eight \gls{spm} banks (\emph{superbanks}) at once through a secondary interconnect\rev{, transferring} up to \SI{64}{\byte} per cycle or \SI{64}{\gibi\byte\per\second}\rev{;}
\rev{this way, it maximizes throughput without significantly slowing down ongoing memory-intensive computations, which can access up to 24 \gls{spm} banks at once using all \glspl{su}.} 
To reduce backpressure in the group- and chiplet-level interconnect, the \gls{dma} has priority accessing the \gls{spm} through the secondary interconnect over the cores and \glspl{su}.
We will elaborate on the \gls{dma} engine's programming model and its use in double-buffered data tiling in~\Cref{sec:progmodel_dm}.

\subsection{Compute Group}
\label{sec:arch_group}

Four clusters together form the next compute hierarchy level, a \emph{group} \blackcircle{3}, shown in \Cref{fig:arch_group}. 
Clusters within a group have full-bandwidth access to each other through two fully connected \gls{axi4} crossbars: a 512-bit crossbar used by \gls{dma} engines and instruction caches for bulk transfers and an atomics-capable 64-bit crossbar used by cores for global synchronization and message passing.
\rev{Thus, groups allow their clusters to locally share data at higher bandwidths than at the global level and constitute a replicable multi-cluster design that significantly simplifies top-level chiplet implementation.}

On each crossbar, a group has one outgoing and one incoming port, providing a shared bandwidth to and from the chiplet interconnect of \SI{64}{\gibi\byte\per\second} for bulk transfers and \SI{8}{\gibi\byte\per\second} for message passing.
\rev{Providing one outgoing port per group best matches the HBM2E bandwidth available at the chiplet level and keeps the chiplet-level interconnect implementable.}
\glspl{iotlb} on the outgoing ports allow for per-group address remapping and access control at page granularity.
A remappable \SI{32}{\kibi\byte} constant cache on the outgoing 512-bit port can be used to cache program code and other immutable data.
Finally, the groups can be individually clock-gated, reset, and isolated from their interconnect ports to the chiplet level through memory-mapped registers for online power management.

\subsection{Occamy Chiplets}
\label{sec:arch_chiplet}

\Cref{fig:arch_chiplet} shows Occamy's top-level chiplet architecture \blackcircle{4}. 
Each chiplet features six groups, totaling 216 cores, and a single Linux-capable 64-bit \riscv~host processor managing the groups and all other on-chip resources.
Like the groups, the chiplets feature a hierarchical 512-bit \gls{axi4} network for bulk data transfers and an atomics-capable 64-bit \gls{axi4} network for synchronization, message passing, and management.
The \gls{d2d} link, which serializes cross-chiplet transactions, comprises a \emph{narrow} and a \emph{wide} segment carrying 64-bit and 512-bit transactions, respectively.

The 512-bit network is composed of three fully connected crossbars.
Each group's \SI{64}{\gibi\byte\per\second} outgoing port provides access to an \emph{HBM} crossbar and a \emph{group} crossbar, interconnecting the six on-chip groups.
The \emph{HBM} crossbar provides access to the eight on-chip \rev{\cn{\SI{47.68}{\gibi\byte\per\second}}} HBM2E controllers (\rev{\cn{\SI{381.47}{\gibi\byte\per\second}}} in total).
It can be configured at runtime to interleave the HBM2E channels at page granularity, facilitating load balancing and data reuse.
A \emph{system} crossbar connects the group and HBM crossbars to the wide D2D segment and 64-bit network, providing all actors across both chiplets access to the entire memory space.
It also connects to a \SI{1}{\mebi\byte} \gls{spm} used for low-latency on-chip storage of shared data and a chiplet-level \gls{dma} engine used by the host for fast, explicit data movement and memory initialization.
\rev{As a chiplet-level memory, the \SI{1}{\mebi\byte} \gls{spm} was dimensioned to be significantly larger than the \SI{128}{\kibi\byte} cluster \glspl{spm}, but small enough to fit into the area available to chiplet-level logic and routing.}

The 64-bit network consists of a single crossbar. It connects the Linux-capable host and the 64-bit ports of the groups to the narrow \gls{d2d} segment, the 512-bit network, and various peripherals.
It also features a \SI{512}{\kibi\byte} \gls{spm} used by the host for management tasks.
The peripherals include UART, I2C, QSPI, GPIOs, a \cn{\SI{1.33}{\giga\bit\per\second}} off-interposer \gls{c2c} link, and a JTAG test access point for live host processor debugging.
They also include RISC-V-compliant timers and platform-level interrupt controllers providing interrupts for both the host processor and all on-chip compute cores.

\rev{The main benefit of our hierarchical crossbar-based interconnect over a ring or mesh topology is its \emph{symmetry}: the memory topology looks identical to all compute cores and clusters, meaning that the architectural bandwidth and latency of accesses to each hierarchy level (cluster, group, chiplet) are \emph{constant}. This greatly simplifies programming, as network performance is homogeneous and code can be written in a \emph{cluster-agnostic} way without sacrificing performance.}
\rev{However, crossbars pose challenges in physical implementation, requiring effort to provision sufficient bandwidth while managing area costs. While this was feasible for our design, larger systems may face scalability limitations. In this case, alternative \gls{noc} topologies such as mesh and torus could be more suitable.}

\rev{The \gls{d2d} link enables seamless communication across chiplets, allowing the system to scale to 432 cores and two HBM2E stacks. This approach improves overall performance while avoiding the yield challenges and high manufacturing costs associated with large monolithic dies~\cite{amd_chiplet_isscc20}. The \gls{d2d} link consists of a narrow segment, optimized for synchronization and message passing, and a wide segment, designed for high-throughput bulk data transfers.}

\begin{figure}[t]
\centering
\includegraphics[width=0.97\linewidth]{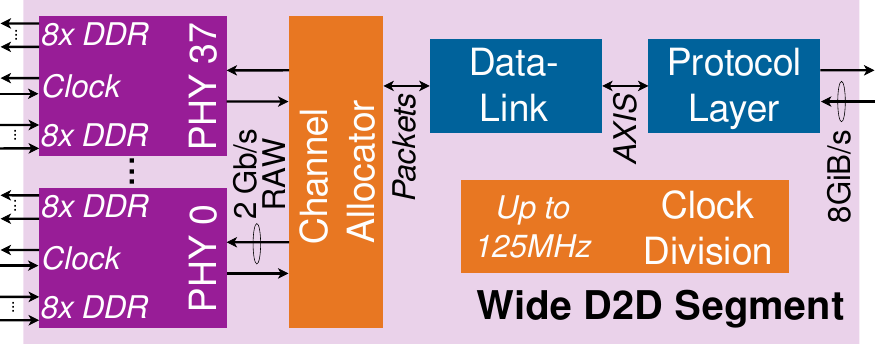}
\caption{Architecture of the wide D2D segment with its 38 source-synchronous double-data-rate PHYs, carrying up to \SI{2}{\gibi\byte\per\second} of raw data at a clock speed of \SI{125}{\mega\hertz}.}
\label{fig:arch_d2d}
\end{figure}

The wide segment is a scaled-up version of the narrow segment with \cn{38} \glspl{phy} to increase bandwidth; its architecture is shown in \Cref{fig:arch_d2d}.
The \emph{protocol layer}, shown in \Cref{fig:arch_d2d}, arbitrates between \gls{axi4} requests and responses and converts them to \gls{axi4}-Stream (AXIS) payloads. 
\rev{It handles the five independent channels of the \gls{axi4} protocol (AW, AR, W, R, B) and applies backpressure to the \gls{axi4} interface to prevent protocol-level deadlocks. Each AXIS payload includes a header that contains information about the packet type and credits required for flow control.}
The \emph{data-link} layer further packetizes the payloads based on the available number of off-chip lanes; 
it also handles credit-based flow-control to ensure that no packets are lost, as well as synchronization and alignment of packets if multiple \glspl{phy} are configured.
\rev{Additionally, the data-link layer features debugging capabilities, including a \emph{Raw Mode} that allows the link to operate independently of the \gls{axi4} interface by sending patterns over specific PHY channels for fault detection.}
Each \gls{phy} features an all-digital and source-synchronous interface with eight \gls{ddr} lanes in each direction.
\rev{On the transmitter side, the PHY operates with a forwarded clock derived from the system clock. On the receiver side, the transmitted packets are synchronized with the system clock and reassembled into the original payload.}
The narrow and wide segments can achieve effective duplex bandwidths of up to \cn{\SI{1.33}{\giga\bit\per\second}} and \SI{64}{\giga\bit\per\second}, respectively.
The wide segment additionally features a \emph{channel allocator} to enable fault tolerance.
An initial calibration detects faulty \glspl{phy}, which can individually be disabled;
the channel allocator then reshuffles packets among the functional \glspl{phy} with only linear bandwidth degradation.
\rev{This fault tolerance mechanism ensures reliable communication across chiplets, even in the presence of manufacturing defects.}

The {\occamy} chiplet relies on \glspl{fll} to generate on-chip clocks for each of its three clock domains: the \emph{compute} domain, the \emph{peripheral} domain, and the \emph{HBM2E PHY} domain. 
The \emph{compute} domain includes the compute groups, the 64-bit host, the chiplet-level interconnect, the \gls{d2d} link, and the HBM2E controller.
Like the groups, the \gls{d2d} link and the HBM2E subsystem can be clock-gated through memory-mapped configuration registers.

\section{Programming Model}
\label{sec:progmodel}

Occamy's compute cores support the unprivileged 32-bit \riscv~\gls{isa} (RV32G) and can be programmed using \riscv~assembly, bare-metal C/C++, or a higher-level code generator, providing the full software agility of a general-purpose system.
An extended LLVM 15 toolchain featuring assembly and intrinsics support for Occamy's \gls{isa} extensions as well as specialized scheduling models and optimization passes allows programmers to write high-performance code targeting dense and sparse applications alike.

Occamy leverages explicit, tiled data movement to achieve near-ideal ($\geq$\SI{75}{\percent}) \gls{fpu} utilizations:
while its \gls{su}-enhanced worker cores process local data tiles in \gls{spm}, the cluster \gls{dma} engines simultaneously transfer double-buffered tiles between the \glspl{spm} and global memory (e.g. HBM2E).
This approach combines the benefits of memory- and interconnect-efficient, latency-tolerant bulk ($\sim$\SI{10}{\kibi\byte}) \gls{dma} transfers and of low-latency, fine-granular access provided to worker cores by the banked \glspl{spm}.

We will first elaborate on how worker cores achieve high cluster-level compute throughput on both dense and sparse applications through the use of \glspl{su}, and then discuss system-level data movement and tiling for full applications.

\definecolor{lstgreen}{rgb}{0.0, 0.59375, 0.0}
\definecolor{lstblue}{rgb}{0.0, 0.0, 0.5}
\definecolor{lstorange}{rgb}{0.55859375, 0.5625, 0.0}
\definecolor{lstred}{rgb}{0.5, 0.0, 0.0}

\begin{figure*}[ht!]
    \begin{subcaptionblock}{0.188\linewidth}%
        \centering%
        \includegraphics[height=18.5em]{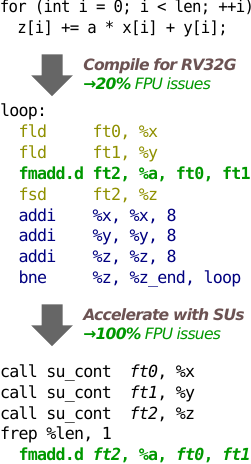}%
        \caption{Dense AXPY}%
        \label{fig:progmodel_axpy}%
    \end{subcaptionblock}\hspace{0.7em}\hfill
    \begin{subcaptionblock}{0.216\linewidth}
        \centering%
        \includegraphics[height=18em]{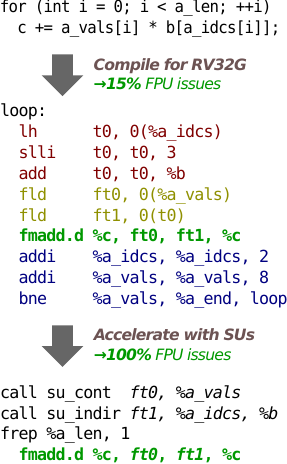}%
        \caption{Sparse-dense DOTP~\cite{scheffler2023sssr}}%
        \label{fig:progmodel_spvv}%
    \end{subcaptionblock}\hfill
    \begin{subcaptionblock}{0.284\linewidth}
        \centering%
        \includegraphics[height=18em]{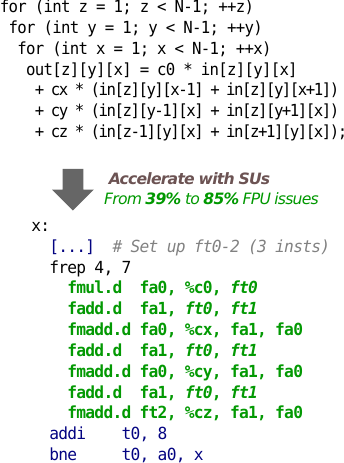}%
            \caption{7-Point stencil code~\cite{scheffler2024saris}}%
        \label{fig:progmodel_stencil}%
    \end{subcaptionblock}\hfill\hspace{-1.0em}
    \begin{subcaptionblock}{0.24\linewidth}
        \centering%
        \includegraphics[height=18em]{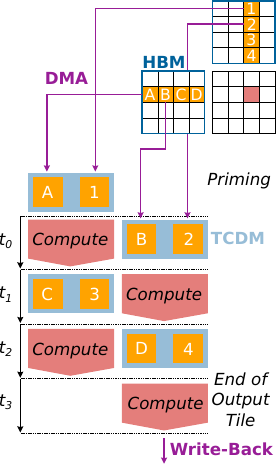}%
        \caption{Data movement scheduling}%
        \label{fig:progmodel_tiling}%
    \end{subcaptionblock}
    \caption{(a-c) Example compute kernels and how they are accelerated on worker cores using \glspl{su} (\textcolor{lstgreen}{green: useful compute}, \textcolor{lstblue}{blue: iteration}, \textcolor{lstorange}{orange: load-stores}, \textcolor{lstred}{red: indirection}) and (d) high-level overview of data movement scheduling for tiled workloads.}
    \label{fig:progmodel}
\end{figure*}

\subsection{SU Performance Benefits}
\label{sec:progmodel_wsu}

Our \glspl{su} maximize \gls{fpu} utilization on both dense and sparse workloads by eliminating the control overheads of streaming \gls{spm} accesses.
They obviate address calculations and load-store issues, hide \gls{spm} access stalls, and enable multiple concurrent load-stores without a superscalar core. 
\rev{In this section, we demonstrate on three simple FP64 compute kernels (\Cref{fig:progmodel_axpy,fig:progmodel_spvv,fig:progmodel_stencil}) how our \glspl{su} can be programmed to leverage the above benefits and enable the significant performance gains we later present in detail in \cref{sec:res_perf}.}

When compiled for the baseline RV32G \gls{isa}, the scaling vector sum (AXPY) in \Cref{fig:progmodel_axpy} issues at most one useful \gls{fma} every eight cycles\rev{:}
\rev{three of the shown loop instructions perform load-stores (shown in orange) and another four handle iteration control (shown in blue), resulting in an eight-instruction loop in which only \emph{one} instruction (the \gls{fma} \texttt{fmadd.d}) utilizes the \gls{fpu} to perform useful compute.}
While code optimizations can further reduce iteration overheads, they cannot eliminate load-stores; even with a 4\x~loop unroll, we observe an \gls{fpu} issue fraction (and thus maximum \gls{fpu} utilization) of only \cn{\SI{20}{\percent}}.
\rev{When accelerating this kernel with \glspl{su}, we eliminate all load-stores and address calculations by merging them into the register accesses of \texttt{fmadd.d}. Before entering the loop, we configure our three \glspl{su}, connected to the registers \texttt{ft0} to \texttt{ft2}, to continuously stream the operands \texttt{x} and \texttt{y} from \gls{spm} and the result \texttt{z} to \gls{spm} as shown. When the useful \texttt{fmadd.d} instruction accesses these registers, the \glspl{su} implicitly perform the associated address calculations and load-stores in hardware.}
Together with our \glsunset{frep}\gls{frep} hardware loop, \rev{which eliminates the loop branch, the loop now consists only of the useful \gls{fma} instruction, which is issued repeatedly as shown.} In practice, kernel setup overheads and \gls{spm} access contention prevent us from achieving \SI{100}{\percent} \gls{fpu} utilization, but \rev{Occamy still reaches an average \gls{fpu} utilization of \SI{78}{\percent} on dense \gls{la} workloads as we will show in \cref{sec:res_perf}}.

\Cref{fig:progmodel_spvv} shows the code for \rev{a} sparse-dense dot product. 
The sparse operand $a$ is stored as two arrays, one holding nonzero values (\texttt{a\_vals}) and another holding their positions (\texttt{a\_idcs}).
In addition to those seen in AXPY, the sparsity-induced \emph{indirection} \texttt{b[a\_idcs[i]]} incurs further address computation overheads on the baseline RV32G \gls{isa} \rev{(shown in red)}, limiting the fraction of useful \gls{fpu} issues to \cn{\SI{15}{\percent}} even with a 4\x~loop unroll.
Our \glspl{su} can \rev{accelerate this kernel by handling} indirection fully in hardware\rev{: the \gls{su} of \texttt{ft1} is configured to fetch the indices \texttt{a\_idcs} and use them to read from the dense vector \texttt{b}, yielding the values \texttt{b[a\_idcs[i]]} when \texttt{ft1} is read. In parallel, the \gls{su} of \texttt{ft0} continuously reads the sparse values \texttt{a\_vals}. By repeatedly issuing \texttt{fmadd.d} using \gls{frep} as shown, we can accumulate the partial products \texttt{a\_vals[i]*b[a\_idcs[i]]} without any non-compute instructions in the loop.}
The \gls{fpu} utilization achieved in practice is no longer bound by control overheads, but by the \gls{spm}'s throughput and stream length imbalances~\cite{scheffler2023sssr}.

\Cref{fig:progmodel_stencil} shows \rev{one time iteration of} a star-shaped stencil computation on a 3D grid. 
Stencils with complex shapes incur irregular memory access patterns, which are hard to optimize and prone to contention.
Through predefined index arrays, our indirection-capable \glspl{su} can stream \emph{arbitrary} \gls{spm} access sequences\rev{, eliminating the associated load-store and address calculation issues. }%
\rev{In the example shown, we can store the offsets of all reads from the \texttt{in} grid relative to the center point \texttt{in[x][y][z]} in constant index arrays; our \glspl{su} can then repeatedly stream these index arrays with a different base address on each iteration of the innermost (\texttt{x}) loop. }
We can leverage this \rev{method, which is described in full detail in~\cite{scheffler2024saris}}, to stream the data for \emph{any} stencil code in ideal processing order. \rev{The} method can \rev{also} be combined with our \gls{frep} hardware loop for grid unrolling as shown.
In our example, the \glspl{su} increase the fraction of FPU issues from \cn{\SI{39}{\percent}} on an RV32G baseline to \cn{\SI{85}{\percent}}.

\begin{figure}
    \begin{subcaptionblock}{0.93375\linewidth}
        \centering%
        \includegraphics[width=\linewidth]{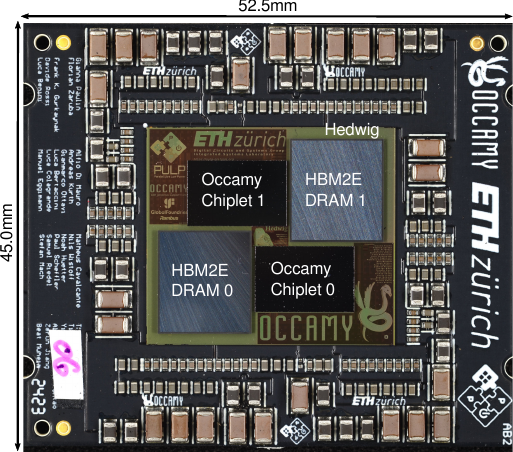}%
        \vspace{-0.5mm}%
        \caption{2.5D Occamy assembly mounted on carrier PCB.}%
        \label{fig:module}%
        \vspace{3.4mm}%
    \end{subcaptionblock}\hfill

    \begin{subcaptionblock}{0.93375\linewidth}
        \centering%
        \includegraphics[width=\linewidth]{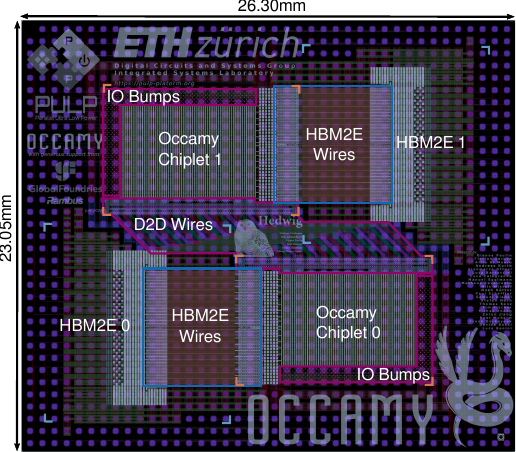}%
        \vspace{-0.5mm}%
        \caption{Layout of the Hedwig interposer.}%
        \label{fig:hedwig:render}%
        \vspace{3.4mm}%
    \end{subcaptionblock}\hfill

    \begin{subcaptionblock}{0.89575\linewidth}
        \hspace{0.35cm}%
        \includegraphics[width=\linewidth]{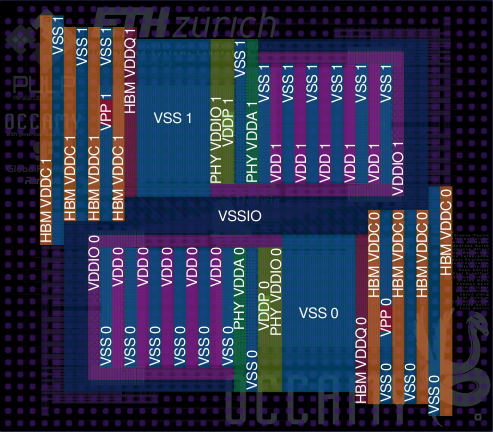}%
        \vspace{-0.5mm}%
        \caption{C4 bump map of the Hedwig interposer.}%
        \label{fig:hedwig:c4}%
    \end{subcaptionblock}\hfill

    \caption{Design of Hedwig interposer and carrier PCB.}
    \label{fig:assembly}
\end{figure}

We re-emphasize that our \glspl{su} are \emph{general-purpose} extensions; they can accelerate \emph{any} data-driven workload involving $\leq$4D affine streams, indirect streams, tensor intersection, or union.
Furthermore, while we focused on \riscv~assembly here to give precise insight into performance at the instruction level, \glspl{su} can be programmed entirely using the dedicated high-level (C/C++) compiler intrinsics we provide in our extended LLVM 15 toolchain.

\subsection{DMA Engines and Data Movement}
\label{sec:progmodel_dm}

Occamy features a programmable \gls{dma} engine in each cluster to facilitate efficient data movement between the local \glspl{spm}, the HBM2E DRAM, and other clusters or chiplets. The \gls{dma} engine optimizes data transfer by managing asynchronous, double-buffered data tiles, ensuring data movement overlaps with computation to maximize resource utilization.

To implement double buffering, the \gls{dma} engine is programmed to fetch the next data tile into a secondary buffer while the worker cores process the current tile. Once the computation on the current tile is complete, the \gls{dma} engine swaps the buffers, making the new data immediately available for processing. 
The \SI{128}{\kibi\byte} cluster \glspl{spm} can support tiles tens of \si{\kibi\byte} in size, which is sufficient to hide the typical access latencies in the hundreds of cycles that accesses to HBM2E or the other chiplet may incur.

\Cref{fig:progmodel_tiling} illustrates our double-buffered tiling strategy. In the diagram, the \emph{compute} blocks represent the phases where the worker cores process data, while the \emph{DMA} blocks correspond to the data transfer phases where new tiles are being transferred into or out of the \gls{spm}.

\section{Silicon Implementation}
\label{sec:impl}

The full Occamy 2.5D system was implemented and fabricated along with a dedicated carrier board, resulting in the compute module shown in \Cref{fig:module}. 
The \cn{\SI{73}{\milli\meter^2}} chiplets were fabricated in \gfs \SI{12}{\nano\meter} LP+ FinFET node using a 13-metal stack.
The two compute dies with their respective Micron \emph{MT54A16G808A00AC-32} HBM2E stacks were mounted on \emph{Hedwig}, a passive, 4-metal-stack \SI{65}{\nano\meter} PKG-25SI interposer from \gf, shown in \Cref{fig:hedwig:render}.

\begin{figure}
    \includegraphics[width=\linewidth]{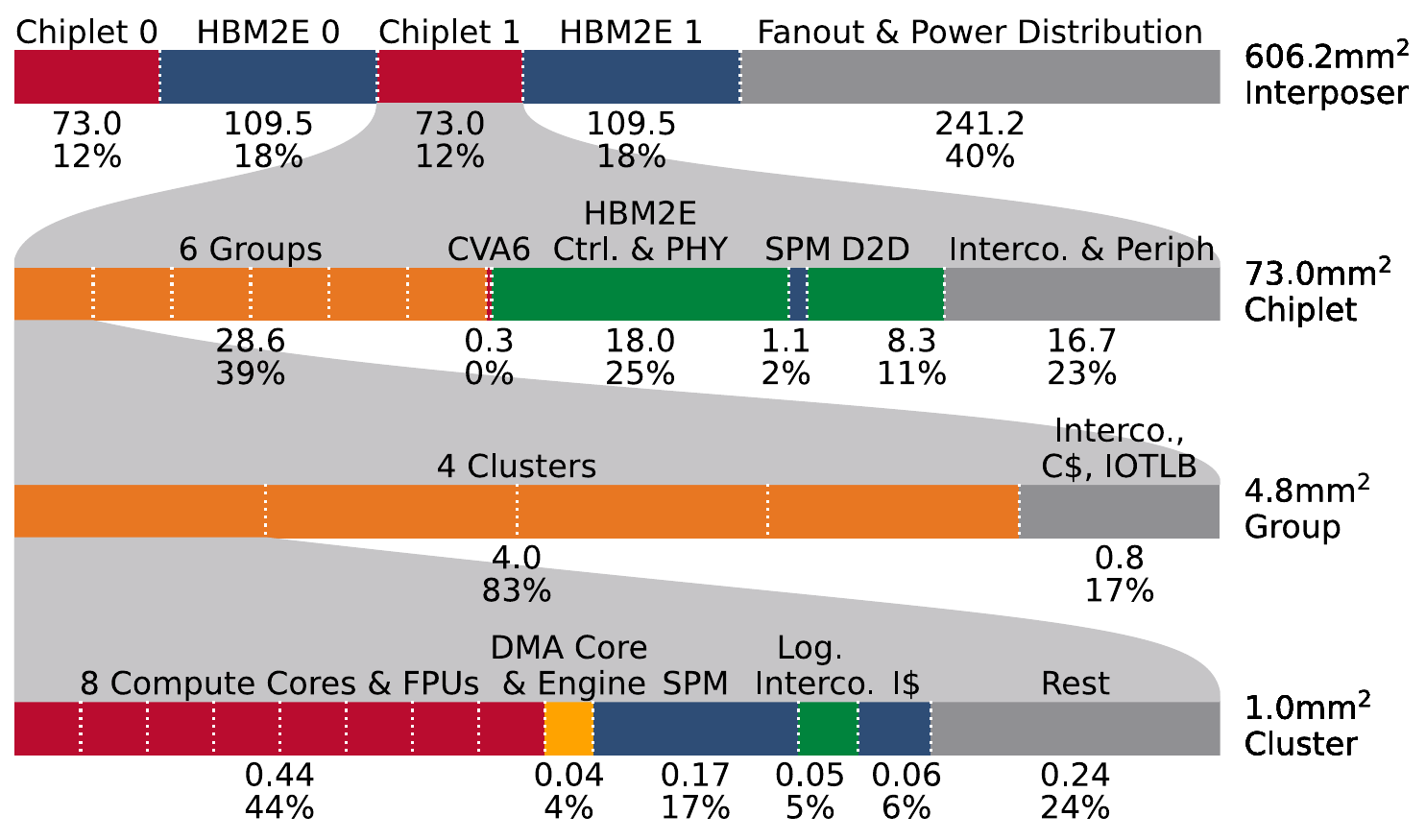}
    \caption{Hierarchical area breakdown of the Occamy system.}
    \label{fig:impl_area}
\end{figure}

\begin{figure*}[ht!]
    \begin{subcaptionblock}{0.277\linewidth}
        \centering%
        \includegraphics[width=\linewidth]{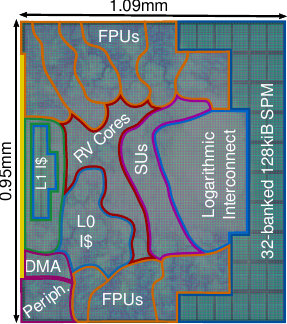}%
        \vspace{-1.3mm}%
        \caption{Compute cluster}%
        \vspace{1.2mm}%
        \label{fig:impl_cluster}%
        \vfill%
        \centering%
        \includegraphics[width=\linewidth]{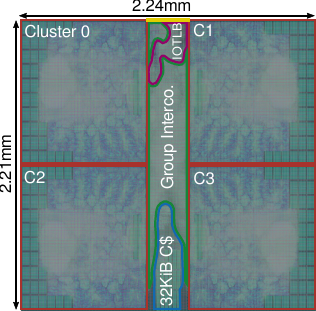}%
        \vspace{-1.3mm}%
        \caption{Compute group}%
        \label{fig:impl_group}%
    \end{subcaptionblock}\hfill
    \begin{subcaptionblock}{0.705\linewidth}
        \centering%
        \includegraphics[width=\linewidth]{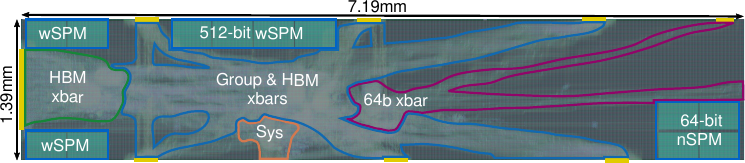}%
        \vspace{-1.3mm}%
        \caption{Chiplet-level interconnect}%
        \vspace{1.2mm}%
        \label{fig:impl_interco}%
        \vfill%
        \centering%
        \centering%
        \includegraphics[width=\linewidth]{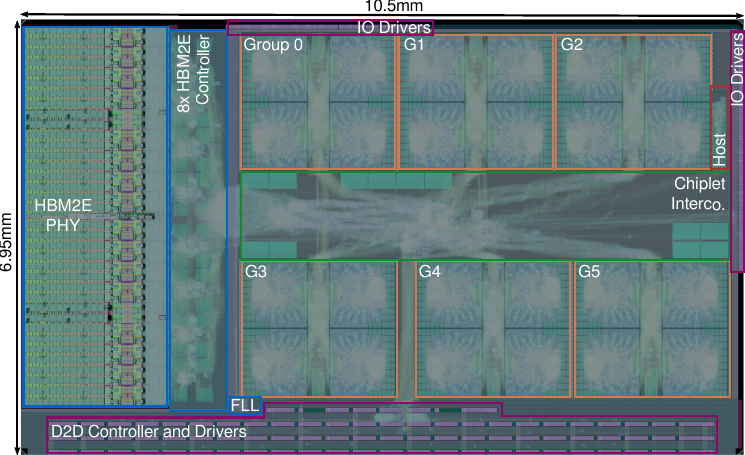}%
        \vspace{-1.3mm}%
        \caption{Chiplet}%
        \label{fig:impl_chiplet}%
    \end{subcaptionblock}\hfill
    \caption{Annotated physical layouts of the {\occamy} chiplet and its hierarchical components (block IO is shown in yellow). }
    \label{fig:impl_chip}
\end{figure*}

\subsection{Occamy Chiplets}

The compute chiplets were synthesized, placed, and routed hierarchically using Synopsys' \emph{Fusion Compiler} 2022.3.
We used 7.5-track standard cell libraries from Arm and HBM2E \glsunset{ip}\glspl{ip} (controller and \gls{phy}) provided by Rambus.
We targeted a nominal compute domain clock of \SI{1}{\giga\hertz} under typical conditions (\SI{0.8}{\volt}, \SI{25}{\celsius}) with a worst-case 
constraint of \SI{950}{\mega\hertz}. 
The peripheral domain was constrained to \SI{500}{\mega\hertz} under worst-case conditions.
The HBM2E controllers and \gls{phy} were constrained as specified to match the DRAM's peak \cn{\SI{3.2}{\giga\bit\per\second\per pin}} throughput.

\rev{\Cref{fig:impl_area} presents a hierarchical area breakdown of an entire {\occamy} assembly.}
\rev{\Cref{fig:impl_chip} shows the resulting hierarchical chiplet layout; we describe our implementation in a bottom-up fashion.}
In accordance with our findings in \cite{paulin2022softtiles}, we arranged the compute cluster's IO ports and L1 instruction cache on one side and its \gls{spm} \glsunset{sram}\glspl{sram} in a U shape on the opposite side, resulting in the \cn{\SI{1.0}{\milli\meter^2}} cluster layout shown in \Cref{fig:impl_cluster}. 
The cluster is area-dominated by the nine RISC-V compute cores with extended FPU functionality (\cn{\SI{44}{\percent}}) and \gls{spm} (\cn{\SI{17}{\percent}}). 
The group layout, shown in \Cref{fig:impl_group}, is almost entirely comprised of its four cluster macros (\cn{\SI{83}{\percent}}) and funnels its shared interconnect ports to a narrow interval on its north edge.
The chiplet interconnect is shown in \Cref{fig:impl_interco}; the global \gls{spm} \glspl{sram} were placed to avoid obstructing the shortest path for crossbar connections.
Further,  we implemented the host processor, the eight HBM2E controllers, and the \gls{d2d} link as hierarchical macros to simplify integration.

The top-level chiplet layout is shown in \Cref{fig:impl_chiplet}; 
it is area-dominated by the six compute groups (\cn{\SI{39}{\percent}}), HBM2E interface (\cn{\SI{25}{\percent}}), and \gls{d2d} link (\cn{\SI{11}{\percent}}).
In accordance with the interposer arrangement, the west and south chiplet beachfronts are fully reserved for the HBM2E and \gls{d2d} interfaces, respectively, while the remaining off-interposer IO drivers are kept on the north and west edges.
The central chiplet bumps are reserved for power delivery.

\subsection{Hedwig Interposer}
\label{sec:hedwig}

The Hedwig interposer is at the center of the 2.5D system, connecting the Occamy chiplets to each other, to their HBM2E stacks, and to the carrier board. 
On its top side, it exposes \cn{\SI{45}{\micro\metre}}-wide \emph{microbump} openings accepting connections from the Occamy chiplets and HBM2E stacks through \cn{\SI{15}{\micro\metre}}-wide copper micropillars. 
On its bottom side, \cn{{1399}} custom-designed hexagonal \emph{C4 pads} with \cn{\SI{320}{\micro\metre}} openings and a \cn{\SI{650}{\micro\metre}} pitch allow us to solder the 2.5D assembly onto the carrier PCB using low-temperature bismuth-based \emph{C4 bumps}.

The connections between the Occamy chiplets and HBM2E stacks were routed on the odd metal layers with a wire width of \cn{\SI{2.5}{\micro\metre}} and a pitch of \cn{\SI{4.1}{\micro\metre}}. 
The HBM2E wire length was kept below \cn{\SI{4.9}{\milli\metre}}, and ground shields were introduced on the even layers to ensure signal integrity. 
\Cref{fig:hedwig:render} highlights the HBM2E wires; the routing density inside the rectangles approaches \cn{\SI{100}{\percent}}. 
The \cn{39} \gls{d2d} link channels were length-matched to \cn{\SI{8.8}{\milli\metre}} and combined into \cn{24} wire bundles located at the center of Hedwig as shown in \Cref{fig:hedwig:render}. 
The bundles were routed on the three upper metal layers in an alternating pattern, leaving the lowest layer for a shared ground connection. 
A routing width of \cn{\SI{3.2}{\micro\metre}} with a pitch of \cn{\SI{6.4}{\micro\metre}} was chosen.

The digital IO of the Occamy chiplets and HBM2E stacks was routed to Hedwig's edge using \cn{\SI{2.5}{\micro\metre}} traces on the three uppermost layers; 
placing IO-related C4 pads at the edge of Hedwig facilitates the fan-out to the carrier PCB.
\Cref{fig:hedwig:c4} shows Hedwig's C4 pad map, highlighting the IO-related connections as well as the sixteen power and three ground domains. 
The ground pads of IO drivers of the two chiplets are connected through the shielding of the \gls{d2d} link, resulting in a shared IO ground \emph{VSSIO}.
For the rest of the power domains, each chiplet-HBM2E pair has its own ground \emph{VSSx}.
The compute fabric, host, and HBM2E controller on each chiplet are supplied through the core power net \emph{VDDx};
the remaining power domains supply the HBM2E PHYs and memory stacks.

\subsection{Carrier PCB}
\label{sec:carrier-pcb}

The carrier PCB completes the Occamy assembly at the lowest level.
It serves three main purposes: giving the assembly mechanical stability, implementing the fan-out of the IO signals, and stabilizing the individual power domains.
The \cn{12}-layer stack-up with \cn{\SI{70}{\micro\metre}}-thick copper foil and \emph{ROGERS RO4350B} high-stability, low-CTE laminate ensures high current delivery capabilities while providing sufficient mechanical strength. 
Decoupling capacitors are placed close to Hedwig on the carrier to comply with the power delivery requirements of the HBM2E stacks.  

The carrier PCB implements an industry-standard \emph{LGA 2011-3} CPU socket interface compatible with off-the-shelf mainboard sockets, facilitating the creation of application boards. 
Most of the \cn{2011} pins are used for power and ground connections, with \cn{877} exposed IOs. 
The carrier is designed to be strong enough to protect the fragile, \cn{\SI{120}{\micro\metre}}-thick Hedwig interposer from mechanical stress while being placed into and removed from the test socket.

\section{Experimental Results}
\label{sec:results}

We create a custom testing infrastructure for {\occamy} as described in \cref{sec:silicon-setup} to evaluate its compute performance and energy efficiency in \cref{sec:res_perf}.
\rev{We further evaluate Occamy's performance and energy efficiency on \gls{ml} inference applications in \cref{sec:res_application}}.
\rev{Finally}, we evaluate the bandwidth scaling and energy efficiency of {\occamy}'s \gls{d2d} link in \Cref{sec:res_die2die}.

\subsection{Silicon Measurement Setup}
\label{sec:silicon-setup}

\begin{figure}
    \begin{subcaptionblock}{0.475\linewidth}
        \centering%
        \includegraphics[width=\linewidth]{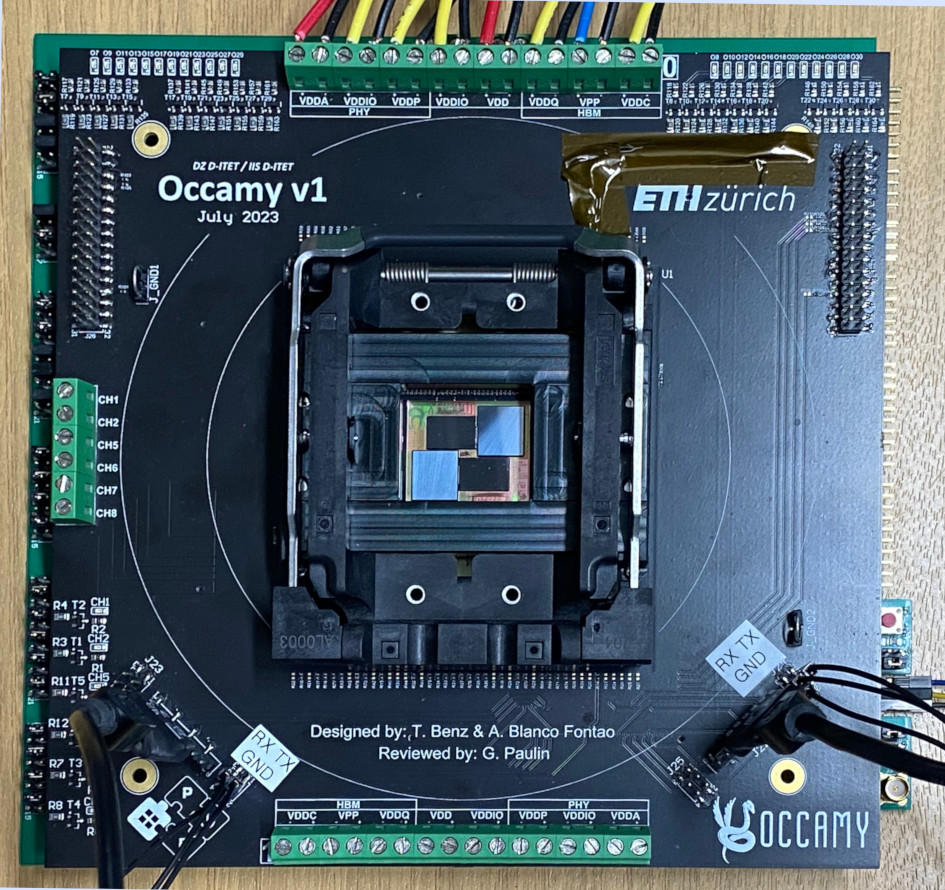}%
        \vspace{-1.7mm}%
        \caption{}%
        \label{fig:tester-pcb}%
    \end{subcaptionblock}\hfill
    \begin{subcaptionblock}{0.49\linewidth}
        \centering%
        \includegraphics[width=\linewidth]{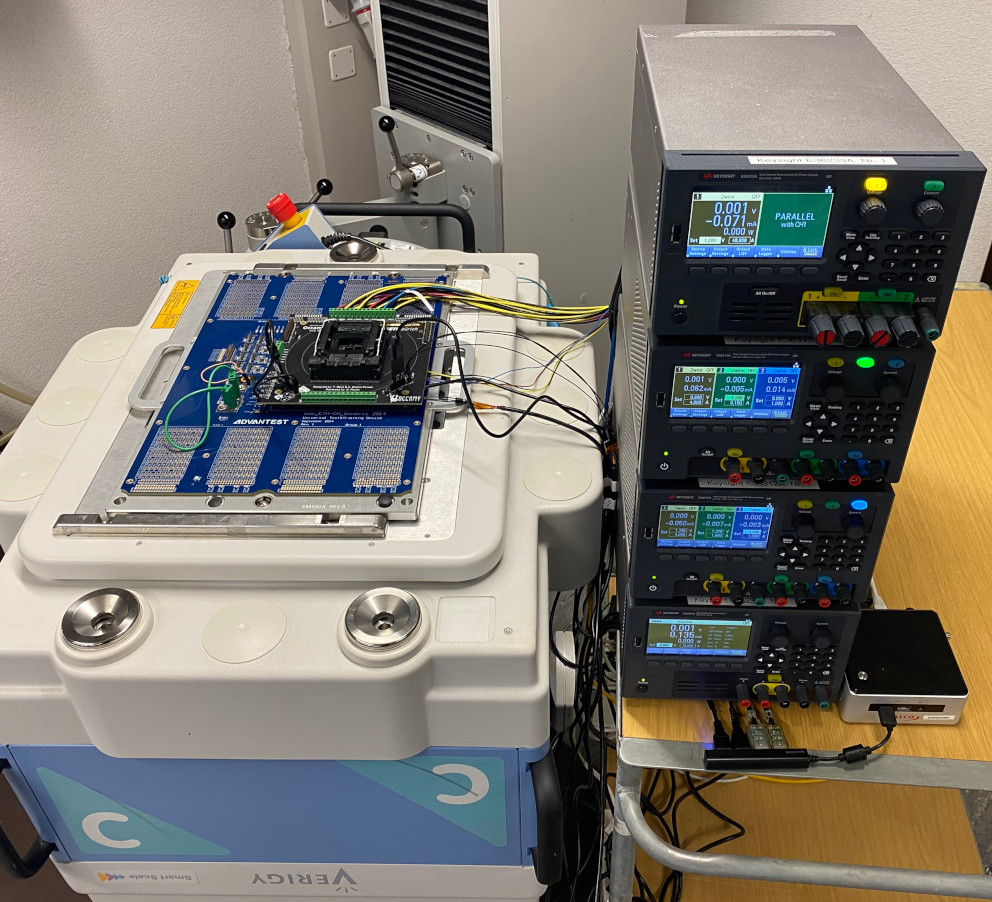}%
        \vspace{-1.7mm}%
        \caption{}%
        \label{fig:setup}%
    \end{subcaptionblock}\hfill
\caption{(a) Bringup board enabling both testing on an %
V93000 ATE and standalone operation, and (b) measurement setup.}
\label{fig:impl_board}
\end{figure}

\Cref{fig:tester-pcb} shows {\occamy}'s \emph{bringup} board, which consists of a stack of two individual PCBs. 
The upper PCB holds an {LGA 2011} \gls{zif} socket for the {\occamy} system, connectors for the \cn{16} power domains, an interface to a \emph{V93000} \gls{ate} system, as well as JTAG, UART, and GPIO headers. 
For standalone operation, the lower PCB provides clocking resources, reset circuitry, configuration headers, and an SD card slot for each  chiplet.

To evaluate {\occamy}, we connect the bringup board to an \gls{ate} system,
ensuring a stable, low-jitter clock delivery. 
In all experiments, the temperature of the {\occamy} system is kept at \cn{25°C} through an active temperature forcing system.
We use Keysight \emph{E36200} series power supplies for power delivery and current measurements.
Two Digilent \emph{JTAG-HS2} programmers are used to program the {\occamy} system through {JTAG}, and two {FTDI} chips relay {UART} data to a testing workstation.
We coordinate the power supplies, resets, programmers, and UART adapters through a dedicated Python script running on our testing workstation, which enables the automated, reproducible execution of applications on Occamy and the collection of current measurements across domains. 

\subsection{Compute Performance and Efficiency}
\label{sec:res_perf}

\begin{figure*}[ht!]
    \hfill
    \begin{subcaptionblock}{0.1982055465\linewidth}
        \centering%
        \includegraphics[width=0.98\linewidth]{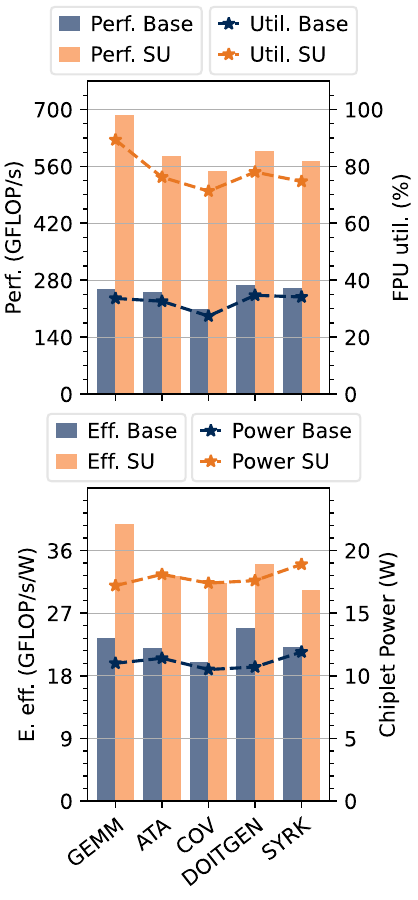}%
        \caption{Dense LA}%
        \label{fig:res_dense}%
    \end{subcaptionblock}\hfill
    \begin{subcaptionblock}{0.3001631321\linewidth}
        \centering%
        \includegraphics[width=0.98\linewidth]{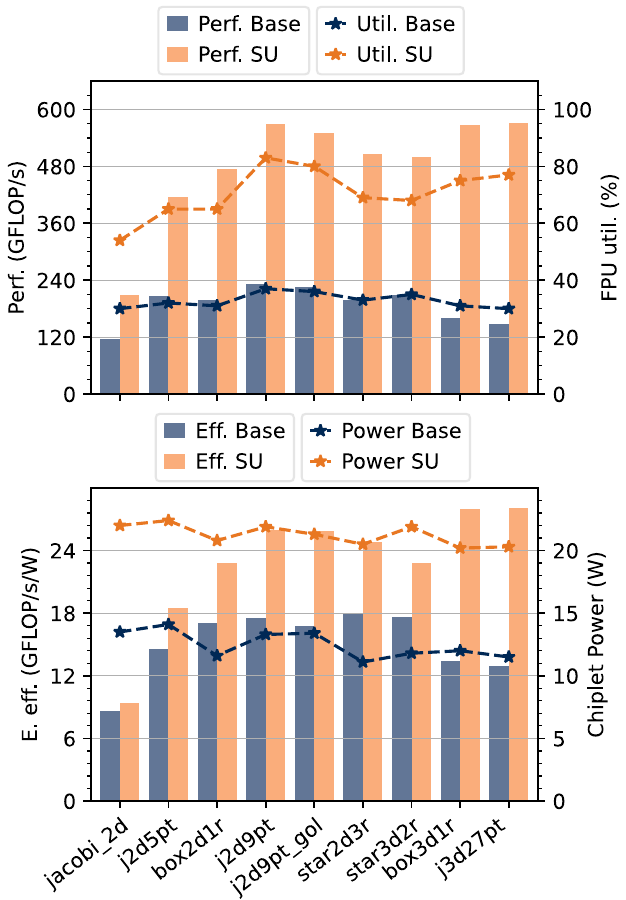}%
        \caption{Stencil codes}%
        \label{fig:res_stencil}%
    \end{subcaptionblock}\hfill
    \begin{subcaptionblock}{0.248776509\linewidth}
        \centering%
        \includegraphics[width=0.98\linewidth]{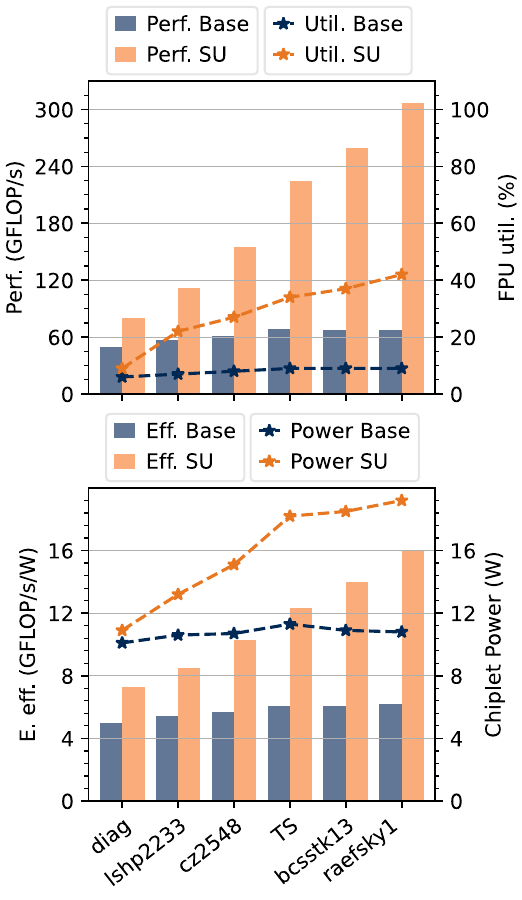}%
        \caption{SpMM}%
        \label{fig:res_spmm}%
    \end{subcaptionblock}\hfill
    \begin{subcaptionblock}{0.248776509\linewidth}
        \centering%
        \includegraphics[width=0.98\linewidth]{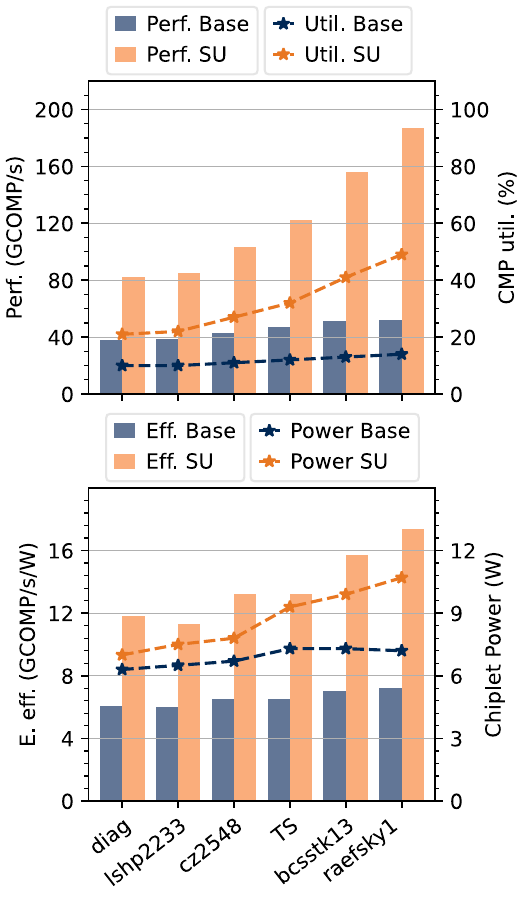}%
        \caption{SpMSpM}%
        \label{fig:res_spmspm}%
    \end{subcaptionblock}\hfill
    \caption{\fontsize{9.925pt}{12.1pt}\selectfont Performance (top), energy efficiency, and power (bottom) for double-buffered FP64 workloads with and without \gls{su} acceleration. SpM(Sp)M left matrices (X axis) are real-world sparse matrices. SpMSpM right matrices are random with \cn{\SI{1}{\percent}} density.}
    \label{fig:res_workloads}
\end{figure*}

\Cref{fig:res_workloads} summarizes {\occamy}'s performance and chiplet energy efficiency at the nominal \SI{1}{\giga\hertz} compute clock on double-buffered dense and sparse FP64 compute workloads with and without \gls{su} acceleration.
We use our ATE-based silicon measurement setup to determine execution times and chiplet power consumption at room temperature, repeating execution as necessary to allow for stable current measurements.
Externally unobservable workload characteristics, such as the number of \glsunset{flop}\glspl{flop} or comparisons performed, are obtained analytically or through
cycle-accurate simulation.

\Cref{fig:res_dense} shows our results on FP64 dense \gls{la} workloads.
Through affine streams, our \glspl{su} accelerate \gls{gemm} using \cn{$48^2$} data tiles by \cn{\SI{2.7}{\x}} over baseline RV32G code, achieving a near-ideal \gls{fpu} utilization of \cn{\SI{89}{\percent}}, a performance of \cn{\SI{686}{\giga\flop\per\second}}, and an energy efficiency of \cn{\SI{39.8}{\giga\flop\per\second\per\watt}}.
On average, \gls{su}-accelerated dense workloads using the same \cn{$48^2$} 
tile size achieve a somewhat lower FPU utilization of \cn{\SI{78}{\percent}} as kernel and \gls{su} setup overheads become more pronounced and fewer FPU operations can be issued as fused (e.g. \gls{fma}) instructions.
Nevertheless, \glspl{su} enable average speedups and energy efficiency improvements over RV32G of \cn{2.4\x}~and \cn{1.5\x}, respectively.

\Cref{fig:res_stencil} shows our results on various FP64 stencil codes scaled out as described in \cite{scheffler2024saris}, using \cn{$64^2$} and \cn{$16^3$} tiles for 2D and 3D grids, respectively.
Our indirect \glspl{su} accelerate all stencil codes by up to \cn{3.9$\times$} compared to heavily optimized RV32G baseline code, achieving a peak FPU utilization of \cn{\SI{83}{\percent}} and up to \cn{\SI{571}{\giga\flop\per\second}} on the \emph{j3d27pt} stencil. 
They also improve peak energy efficiency from \cn{\SIrange[]{17.9}{28.1}{\giga\flop\per\second\per\watt}}.
The remaining execution overheads preventing full \gls{fpu} utilization with \glspl{su} are due to kernel initialization, \gls{su} setup, and \gls{spm} access contentions between worker cores.

\Cref{fig:res_spmm,fig:res_spmspm} show our results for sparse-dense \glsunset{spmm}(SpMM) and \gls{spmspm}, respectively.
In both cases, we multiply real-world left sparse matrices from \cite{Davis2011TheUO}~with random right matrices (\cn{\SI{1}{\percent}} density for \gls{spmspm}); the left matrices are dynamically tiled and reused within compute groups.
\rev{To demonstrate Occamy's flexibility in handling sparse operands, the real-world left matrices are unstructured, cover different application domains, and span from \SI{0.12}{\percent} to \SI{2.8}{\percent} density (sorted left-to-right in figures).}
Through indirection, our \glspl{su} accelerate \gls{spmm} by up to \cn{4.6$\times$}, achieving up to \cn{\SI{307}{\giga\flop\per\second}}, \cn{\SI{16.0}{\giga\flop\per\second\per\watt}}, and \cn{\SI{42}{\percent}} \gls{fpu} utilization.
We introduce index comparison rate as a figure of merit for sparse-sparse matrix computation performance;
through index intersection, our \glspl{su} accelerate \gls{spmspm} by up to \cn{3.6$\times$}, reaching up to \cn{\SI{187}{\giga\comp\per\second}}, \cn{\SI{17.4}{\giga\comp\per\second\per\watt}}, and index comparator utilizations of up to \cn{\SI{49}{\percent}}.
Normalizing our peak performance to the chiplet area dedicated to compute, we achieve up to \cn{\SI{3.63}{\giga\comp\per\second\per\milli\meter^2}}.
For both \gls{spmm} and \gls{spmspm}, performance improves with increasing left matrix density, which helps amortize \gls{su} setup costs as described in~\cite{scheffler2023sssr}.
\rev{Accordingly, while Occamy maintains consistently high utilization as operands become denser, its performance and energy efficiency on extremely sparse operands eventually becomes limited by \gls{su} setup costs and the cluster \glspl{spm}' finite indexing range.}

\Cref{fig:ires_multiprec} shows how \gls{su}-accelerated \gls{gemm} performance and energy efficiency scale as we reduce floating-point precision from 64 to 8 bits, leveraging our FPU's \gls{simd} capabilities.
We consider FP16 GEMM without and with expanding FP32 accumulation (EXP suffix) and FP8 GEMM only with expanding FP16 accumulation.
As we reduce precision, the achieved FPU utilization drops from \cn{\SI{89}{\percent}} (FP64) to \cn{\SI{66}{\percent}} (FP8 EXP).
This is because the SIMD and mixed-precision kernels incur additional overheads: they must zero-initialize accumulators, reduce multiple accumulators packed in a $64$-bit register, and possibly convert the higher-precision result to the original precision~\cite{bertaccini2024fpu}.
The FP64 kernel does not require conversions and avoids accumulator initialization through the non-accumulating \texttt{fmul.d} instruction in the first loop iteration.
With these overheads in mind, the throughput and energy efficiency scale as expected;
for FP8 GEMM with expanding accumulation, we achieve a throughput of \cn{\SI{4.1}{QP\text{-}\tera\flop\per\second}} and an energy efficiency of \cn{\SI{263}{QP\text{-}\giga\flop\per\second\per\watt}}.
On FP16, expanding GEMM is \cn{\SI{6.5}{\percent}}~more energy-efficient than non-expanding GEMM thanks to dedicated expanding dot product units in our FPU~\cite{bertaccini2024fpu}.

\subsection{\rev{ML Inference Applications}}
\label{sec:res_application}
\begin{revenv}

We demonstrate Occamy's architectural benefits on end-to-end applications by evaluating its performance and energy efficiency on two \gls{ml} inference tasks%
: a \gls{gcn} layer, which requires sparse-dense graph aggregation, and a full \gls{gpt}-J \gls{llm}.

To evaluate \glspl{gcn}, we implement the layer computation described in~\cite{kipf2017semisupervised}, which is used in classification tasks and involves both dense (feature recombination) and sparse-dense (graph aggregation) computation steps.
We model the hidden layers of the \gls{gcn} benchmarked in~\cite{dwivedi2024} on node classification, mapping an FP64 layer with 144 input and output features per graph node across both chiplets. We double-buffer incoming features and distribute the layer's output features among the 48 compute clusters.
We measure Occamy's performance and energy efficiency with and without \glspl{su} on the highly sparse citation graphs \emph{webkb}~\cite{lu2003linkbased}, \emph{cora}~\cite{sen2008collective}, and \emph{citeseer}~\cite{sen2008collective}  with average node degrees of 1.8, 2.0, and 1.4, respectively.

\begin{figure}
    \begin{subcaptionblock}{0.49\linewidth}
        \centering%
        \includegraphics[width=\linewidth]{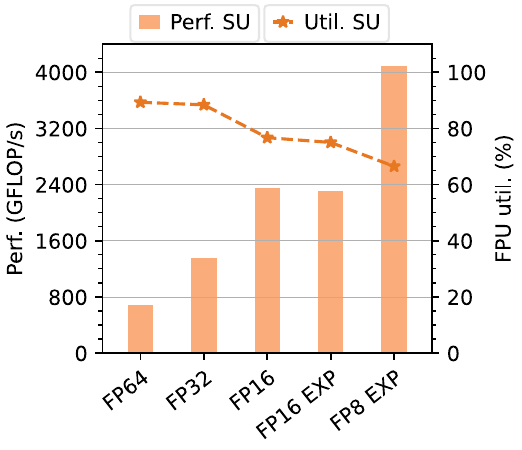}%
        \caption{Performance}%
        \label{fig:ires_multiprec:perf}%
    \end{subcaptionblock}\hfill
    \begin{subcaptionblock}{0.49\linewidth}
        \centering%
        \includegraphics[width=\linewidth]{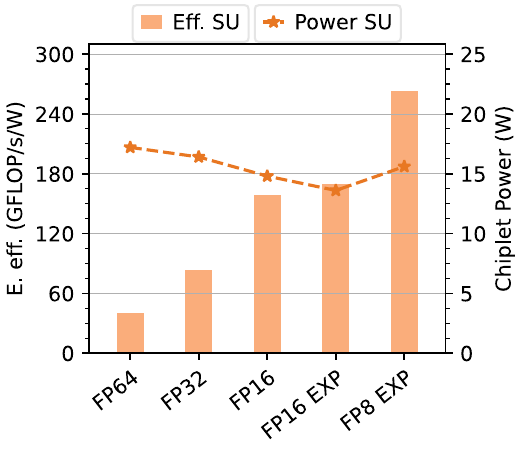}%
        \caption{Energy eff. and power}%
        \label{fig:ires_multiprec:eff}%
    \end{subcaptionblock}\hfill
\caption{Performance, energy efficiency, and power of GEMM at FP64 to FP8 precisions (EXP: expanding accumulation).}
\label{fig:ires_multiprec}
\end{figure}

\Cref{fig:res_gcn} shows the performance, energy efficiency, and power Occamy achieves on  \gls{gcn} layer computation, with the evaluated graphs sorted left-to-right by increasing node count.
Our \glspl{su} enable an average speedup of 2.2\x~over an optimized RV32G baseline, raising the peak \gls{fpu} utilization from \SI{24}{\percent} to \SI{54}{\percent} and achieving up to \SI{413}{DP\text{-}\giga\flop\per\second} or \SI{8.01}{DP\text{-}\giga\flop\per\second\per\milli\meter^2} on \emph{citeseer}.
Peak energy efficiency improves from 16.4 to \SI{25.0}{DP\text{-}\giga\flop\per\second\per\watt}.
As would be expected, our results fall between those achieved for dense and sparse workloads, showcasing Occamy's ability to efficiently handle mixtures of sparse and dense compute.
Overall, we observe a slight increase in \gls{su} performance and energy efficiency as graphs get larger, as this enables better amortization \gls{su} setup costs as was the case for SpMM.

We evaluate \gls{su}-accelerated \gls{llm} inference on Occamy by implementing a full FP16 \gls{gpt}-J model.
Building on the approach in~\cite{potocnik2024optimizing}, we hide communication overheads by staging input and intermediate data transfers concurrently with kernel execution.
In the \acrshort{gpt}-J blocks, both the layer normalization and linear layers are tiled to evenly distribute the workload across all clusters and maximize \acrshort{tcdm} utilization.
The attention mechanism is implemented using the \emph{FlashAttention-2}~\cite{dao2023flashattention} algorithm and fused with the subsequent concatenation and linear projection.
Each chiplet produces half of the final \emph{MHA} layer’s output matrix; the remaining computations, consisting of fused linear and layer normalization layers, are also tiled among the clusters as mentioned before.

\begin{figure}
    \begin{subcaptionblock}{0.49\linewidth}
        \centering%
        \includegraphics[width=\linewidth]{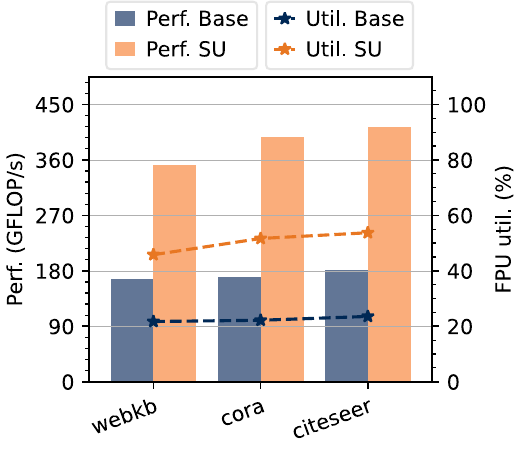}%
        \caption{\rev{Performance}}%
        \label{fig:res_gcn_perf}%
    \end{subcaptionblock}\hspace{0mm}
    \begin{subcaptionblock}{0.49\linewidth}
        \centering%
        \includegraphics[width=\linewidth]{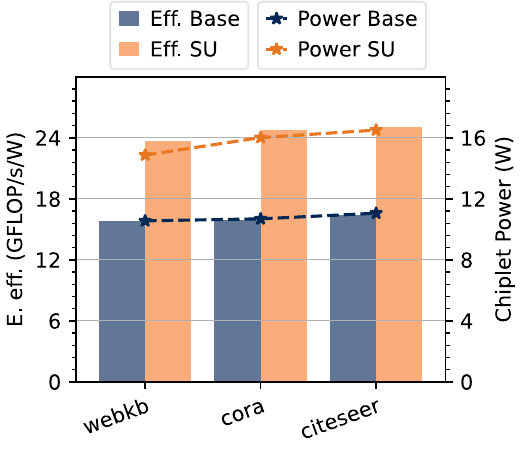}%
        \caption{\rev{Energy eff. and power}}%
        \label{fig:res_gcn_pow}%
    \end{subcaptionblock}\hfill
\caption{\rev{Performance, energy efficiency, and power on \gls{gcn} layer (144\x144, FP64) with and without \gls{su} acceleration.}}
\label{fig:res_gcn}
\end{figure}

\begin{figure}
    \begin{subcaptionblock}{0.44\linewidth}
        \centering%
        \includegraphics[width=\linewidth]{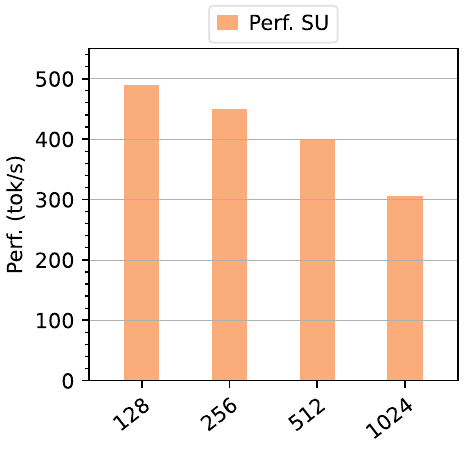}%
        \caption{\rev{Performance}}%
        \label{fig:res_llm_perf}%
    \end{subcaptionblock}\hspace{3mm}
    \begin{subcaptionblock}{0.50\linewidth}
        \centering%
        \includegraphics[width=\linewidth]{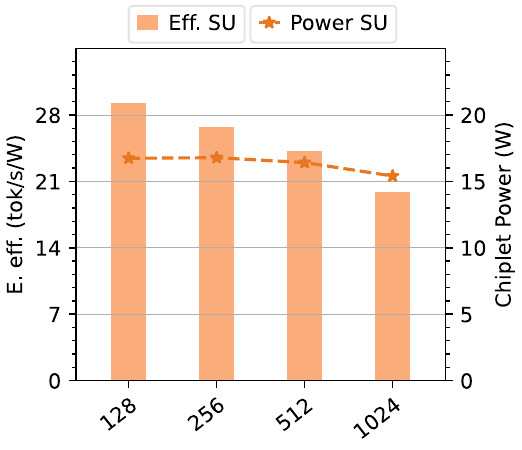}%
        \caption{\rev{Energy eff. and power}}%
        \label{fig:res_llm_pow}%
    \end{subcaptionblock}\hfill
\caption{\rev{Performance, energy efficiency, and power of \gls{su}-accelerated FP16 GPT-J LLM inference in non-autoregressive mode and for varying sequence lengths.}}
\label{fig:res_llm}
\end{figure}

\Cref{fig:res_llm} shows the performance, energy efficiency, and power of Occamy running the full \gls{gpt}-J model in non-autoregressive mode for sequence lengths from 128 to 1024 tokens.
We use the standard measure of \emph{token rate} (\si{\tok\per\second}) to express \gls{llm} performance.
At the shortest sequence length, our implementation achieves a peak throughput of \SI{490}{\tok\per\second} and an energy efficiency of \SI{29.3}{\tok\per\second\per\watt}, achieving a peak FPU utilization of \SI{75}{\percent}.
As can be seen in \Cref{fig:res_llm_perf}, increasing the sequence length leads to a gradual decrease in throughput,
which is due to the well-understood quadratic scaling of attention computation with sequence length.
Accordingly, we see a gradual decrease in energy efficiency with increasing sequence length.
While short sequences are dominated by matrix multiplication, as sequence length grows, more time is spent on the attention mechanism’s less efficient element-wise \emph{softmax} operation.

\end{revenv}

\subsection{Die-to-Die Link}
\label{sec:res_die2die}

\begin{figure}
    \begin{subcaptionblock}{0.49\linewidth}
        \centering%
        \includegraphics[width=\linewidth]{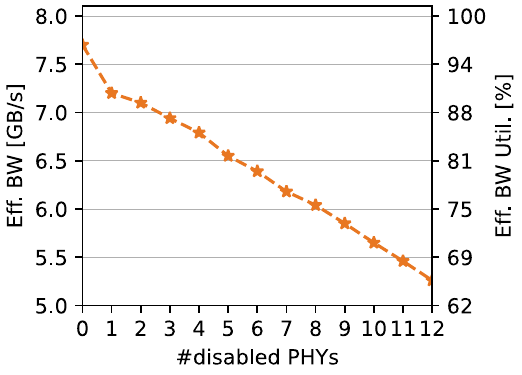}%
        \caption{}%
        \label{fig:d2d_bw:dis_phy}%
    \end{subcaptionblock}\hfill
    \begin{subcaptionblock}{0.49\linewidth}
        \centering%
        \includegraphics[width=\linewidth]{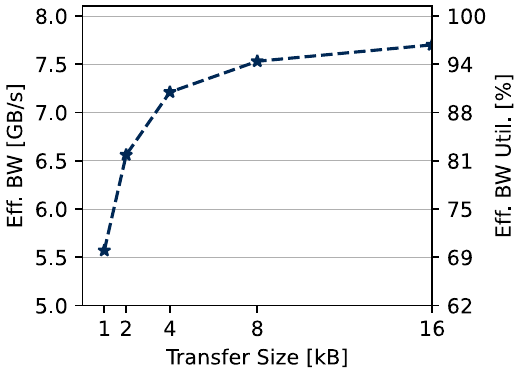}%
        \caption{}%
        \label{fig:d2d_bw:tf_size}%
    \end{subcaptionblock}\hfill
    \caption{(a) Linear effective bandwidth decrease for channel disabling on wide \gls{d2d} segment and (b) effective bandwidth utilization for varying transfer sizes on wide \gls{d2d} segment.}
    \label{fig:d2d_bw}
\end{figure}

\Cref{fig:d2d_bw} shows the effective bandwidth of the wide \gls{d2d} link under varying conditions analyzed in cycle-accurate \gls{rtl} simulation.
As expected, our bandwidth decreases linearly as we increase the number of disabled \glspl{phy} to mitigate faults.
Furthermore, a high link utilization of \cn{96\%} is achieved at a transfer size of \cn{\SI{16}{\kilo\byte}}. 
We also determine the latency of communication between the two chips over the \gls{d2d} link:
accessing the other chiplet's narrow \gls{spm} through the narrow segment from the host core incurs a latency of \cn{27} cycles, while a \gls{dma} transfer over the wide segment experiences a latency of \cn{61} cycles.
On our silicon demonstrator, we measure the energy efficiency of the wide \gls{d2d} segment to be \cn{\SI{1.6}{\pico\joule\per\bit}}.

\section{SoA \rev{Review} \& Comparison}
\label{sec:soacomp}

\rev{We will first review \gls{soa} hardware approaches to sparse compute acceleration, contrasting them with Occamy (\cref{subsec:soadisc}), and then quantitatively compare Occamy's performance and energy efficiency to those achieved on \gls{soa}, silicon-proven CPUs and GPUs (\cref{subsec:soacomp}).}

\subsection{\rev{Sparse Hardware Acceleration SoA}}
\label{subsec:soadisc}

Large-scale \gls{ml} and \gls{hpc} applications primarily target general-purpose CPUs and GPUs.
These architectures efficiently handle dense compute~\cite{nvidia_a100_dense, siegmann2024first}, but achieve low FPU utilization on sparse workloads (\cn{$\leq$\SI{10}{\percent}} on sparse \gls{la}~\cite{alappat2020a64fx, nvidia_a100_sparse, siegmann2024first}) due to their data-dependent control flow and complex, indirect address computations.

To improve sparse compute efficiency, numerous accelerators %
have been proposed.
Most notably, sparse \gls{ml} accelerators achieving leading throughput and energy efficiency on \gls{ml} inference tasks have been demonstrated in silicon~\cite{huang2025hybr, yue2024cimblock, feng2024pntcloud, zhang2021snap};
however, they impose low-precision data formats ($\leq$\SI{16}{\bit}) and low ($\leq$\SI{80}{\percent}) or structured sparsity specific to \gls{ml} workloads and are thus unsuitable for \gls{hpc}.
Generic sparse~\gls{la}~\cite{Srivastava2020MatRaptorAS, Pal2018OuterSPACEAO, park2020output, sadi2019efficient} or tensor algebra~\cite{Hegde2019ExTensorAA} accelerators are less domain-specific, with some demonstrated in silicon~\cite{park2020output, sadi2019efficient}, but most are still hardwired for specific \gls{la} operations~\cite{Srivastava2020MatRaptorAS, Pal2018OuterSPACEAO, park2020output, sadi2019efficient} or dataflows~\cite{Srivastava2020MatRaptorAS, Pal2018OuterSPACEAO}.

Discrete accelerators, while highly efficient, cannot match the flexibility and agility of general-purpose architectures.
Recently, \glsreset{cgra}\glspl{cgra} architectures have been proposed, supporting streaming indirection, intersection, and union in hardware~\cite{Dadu2019TowardsGP,koul2024onyx}. One such \gls{cgra}, \emph{Onyx}~\cite{koul2024onyx}, was recently demonstrated in silicon.
However, \glspl{cgra} are still highly challenging to productively program~\cite{liu2019cgrasurvey}, and Onyx only supports INT16 and BF16 data formats of insufficient precision for most \gls{hpc} workloads.

Another common approach to maintain generality is through \gls{isa} extensions for CPUs and GPUs that accelerate sparse compute~\cite{siracusa2023tmu, Domingos2021UnlimitedVE, zhu2019sparsetensorcore, Wang2019StreambasedMA, Wang2021DualsideST, Rao2022SparseCoreSI};
however, most of these extensions are not demonstrated in silicon~\cite{siracusa2023tmu, Domingos2021UnlimitedVE, zhu2019sparsetensorcore, Wang2019StreambasedMA, Wang2021DualsideST, Rao2022SparseCoreSI} and only accelerate one-~\cite{siracusa2023tmu, Domingos2021UnlimitedVE, zhu2019sparsetensorcore, Wang2019StreambasedMA} or two-sided~\cite{Wang2021DualsideST, Rao2022SparseCoreSI} sparse operations.
Thus, Occamy is the first silicon-proven general-purpose system to efficiently and comprehensively handle sparse workloads in 8-to-64-bit precisions.

\subsection{\rev{Comparison to SoA CPUs \& GPUs}}%
\label{subsec:soacomp}%

\cref{tab:soa} compares Occamy to \gls{soa}, silicon-proven GPUs and CPUs supporting the dominant \glspl{isa} (x86, Arm, \riscv).
To provide a holistic overview, we compare Occamy to both recently released processors (e.g. AMD Genoa, Nvidia H100), which usually lack published stencil and sparse-dense \gls{la} performance evaluations, as well as established processors from prior generations (e.g. Fujitsu A64FX, Nvidia A100) whose real-world performance is well-studied and evaluated in literature.
Occamy's absolute performance is far ahead of other academic \riscv~silicon prototypes such as~\cite{schmidt2022rsoc}, achieving \cn{8.3\x}~higher peak FP64 compute;
it is much closer in peak throughput to the commercial Sophon SG2042~\cite{sophon-trm, chen2020xuantie} \riscv~manycore (\cn{$0.77$} vs \cn{\SI{1.0}{DP\text{-}\giga\flop\per\second}}).
Our %
\gls{d2d} link, while comparatively low in throughput due to its simple \gls{phy} design, achieves a competitive energy efficiency \cn{\SI{23}{\percent}} higher than that in AMD's Rome~\cite{amd_zen2_hc31}.
Unlike other designs in \Cref{tab:soa}, Occamy's entire digital compute architecture, including the groups, interconnect, host, and \gls{d2d} link, is available as a free and open-source SystemVerilog \gls{rtl} description.%

On FP64 dense \gls{la}, Occamy's \glspl{su} enable competitive FPU utilization through affine streams.
Our \cn{\SI{89}{\percent}} peak \gls{fpu} utilization is notably higher than that achieved on AMD's Rome~\cite{a64fx_dense} and Fujitsu's A64FX~\cite{siegmann2024first} and is comparable to the \cn{\SI{95}{\percent}} reached on Nvidia's A100 GPU~\cite{nvidia_a100_dense}.
At up to \cn{\SI{13.3}{DP\text{-}\giga\flop\per\second\per\milli\meter^2}}, dense \gls{la} on Occamy is highly area-efficient:
scaled to our \SI{12}{\nano\meter} node, we achieve a higher compute density than all competitors except Schmidt et al.~\cite{schmidt2022rsoc}, A100, and H100~\cite{hopper2022_hotchips, hopper2024_whitepaper}.
Occamy's peak dense \gls{la} energy efficiency of \cn{\SI{39.8}{DP\text{-}\giga\flop\per\second\per\watt}} is notably higher than that of the \glspl{cpu} we compare to, but cannot reach the efficiency of Nvidia's highly optimized A100 and H100 GPUs fabricated in more advanced nodes.

For FP64 stencil codes, Occamy's indirect \glspl{su} enable significant utilization benefits over the compared CPU and GPU designs. 
Our \cn{\SI{83}{\percent}} peak FPU utilization is \cn{1.7\x}~higher than the \rev{leading} \cn{\SI{49}{\percent}} achieved on A100~\cite{nvidia_a100_stencil}
\rev{using CUDA compute only (to ensure a fair comparison, we normalized this FPU utilization to A100's reduced \SI{9.7}{DP\text{-}\tera\flop\per\second} peak \emph{excluding} tensor cores). While specialized transforms \emph{can} leverage A100's tensor cores to reach a higher peak throughput%
~\cite{chen2024convstencil} (albeit at lower FPU utilization of \SI{31}{\percent}), the corresponding node-normalized compute density %
is still 1.2\x~lower than the \cn{\SI{11.1}{DP\text{-}\giga\flop\per\second\per\milli\meter^2}} Occamy achieves.}
\newcommand{\ct}[1]{\todo{~[#1]}} %
\newcommand{\bs}[1]{{#1}} %
\newcommand{\sh}[1]{{#1}} %
\newcommand{\vf}[1]{{#1}} %
\newcommand{\tn}[1]{\,\textcolor{black}{\tnote{#1}}}
\newcommand{\im}[1]{\item[#1]}
\newcommand{\dk}{\todo{-}} %
\newcommand{\sep}{$\vert$}
\newcommand{\rot}[1]{\rotatebox[origin=c]{90}{#1}}

\renewcommand\cellset{\renewcommand\arraystretch{1.0}%
    \setlength\extrarowheight{0.5ex}}

\def\hdls{1mm}

\afterpage{%
\begin{landscape}
\centering
\begin{table}[p]
\caption{Comparison of Occamy to \gls{soa}, silicon-proven GPUs and CPUs supporting the dominant ISAs (x86, Arm, and RISC-V).}
\label{tab:soa}
\centering
\resizebox{0.85\paperheight}{!}{%
\begin{threeparttable}
\renewcommand{\arraystretch}{1.0}
\begin{tabular}{@{}lllllllllll@{}}
\toprule
& %
& %
\makecell[cl]{\textbf{Schmidt} \\ \textbf{et al.}~\cite{schmidt2022rsoc}} &
\makecell[cl]{\textbf{Sophon}\\  \textbf{SG2042}~\cite{sophon-trm}} &
\makecell[cl]{\textbf{AMD Rome} \\ \cite{amd_zen_2_isscc20, amd_chiplet_isscc20}} & %
\makecell[cl]{\textbf{AMD Genoa} \\ \cite{munger2023zen4isscc, ravi2024zen4micro}} &
\makecell[cl]{\textbf{Fujitsu} \\ \textbf{A64FX}~\cite{a64fx_isscc22}} & %
\makecell[cl]{\textbf{Nvidia} \\ \textbf{A100}~\cite{a100_isscc21}} & %
\makecell[cl]{\textbf{Nvidia} \textbf{H100}\\\cite{hopper2022_hotchips, hopper2024_whitepaper}} &
\makecell[cl]{\textbf{Intel Sapphire} \\ \textbf{Rapids}~\cite{sapphire_rapids_isscc22, sapphire_rapids_hc33}} &
\makecell[cl]{\textbf{\textit{{\occamy}}} \\ \textbf{\textit{(Ours)}}} \\
\midrule\noalign{\vskip \hdls}
\multirow{10}{*}{\rot{\makecell[cc]{\\ General \\}}} &
Assembly &
Single Die &
Single Die &
Organic Substrate &
Organic Substrate &
2.5D Passive Interp. &
2.5D CoWoS\tn{a} &
2.5D CoWoS\tn{a} &
2.5D EMIB\tn{b} &
2.5D Passive Interp. \\
&
\makecell[cl]{2.5D Compute Cfg.} &
\makecell[cl]{\dk} &
\makecell[cl]{\dk} &
\makecell[cl]{8 Die} &
\makecell[cl]{12 Die} &
\makecell[cl]{1 Die} &
\makecell[cl]{1 Die} &
\makecell[cl]{1 Die} &
\makecell[cl]{4 Die} &
\makecell[cl]{2 Die} \\
&
\makecell[cl]{2.5D DRAM Cfg.} &
\makecell[cl]{\dk} &
\makecell[cl]{\dk} &
\makecell[cl]{1 IO Die} &
\makecell[cl]{1 IO Die} &
\makecell[cl]{4 HBM2} &
\makecell[cl]{6 HBM2} &
\makecell[cl]{5 HBM3} &
\makecell[cl]{4 HBM2E} &
\makecell[cl]{2 HBM2E} \\
&
\makecell[cl]{Die Technology} &
\makecell[cl]{16nm FinFET} &
\makecell[cl]{12nm FinFET\tn{c}} &
\makecell[cl]{7nm FinFET} &
\makecell[cl]{5nm FinFET} &
\makecell[cl]{7nm FinFET} &
\makecell[cl]{7nm FinFET} &
\makecell[cl]{TSMC 4N FinFET} &
\makecell[cl]{Intel 7 SuperFin} &
\makecell[cl]{12nm FinFET\tn{d}} \\

&
ISA &
RV64GXhwacha4 &
\makecell[cl]{RV64GCV\\(RVV v0.7.1)} &
x86-64, AVX2 &
x86-64, AVX-512 &
Armv8.2-A SVE\tn{e} &
PTX ISA, CC 8.0 &
PTX ISA, CC 9.0 &
\makecell[cl]{x86-64, AVX-512,\\DL Boost, AMX} &
RV32GXoc\tn{f,g} \\
&
Open-Source RTL &
Partially (Core) &
Partially (Core)\tn{h} &
No &
No &
No &
No &
No &
No &
Partially (Compute) \\
&
\makecell[cl]{%
Freq.\,[\si{\giga\hertz}] \\ Voltage\,[\si{V}]
} &
\makecell[cl]{%
1.44 \\ 1.0
} &
\makecell[cl]{%
2.00 \\ 0.8  
} &
\makecell[cl]{%
\vf{2.25}\tn{p} \\ \vf{0.8}  
} &
\makecell[cl]{%
\vf{2.40}\tn{p} \\ \dk  
} &
\makecell[cl]{%
\vf{2.20} \\ \vf{0.8}  
} &
\makecell[cl]{%
\vf{1.41} \\ \vf{0.8}
} &
\makecell[cl]{%
1.98 \\ \dk  
} &
\makecell[cl]{%
2.60 \\ \dk  
} &
\makecell[cl]{%
\vf{1.00} \\ \vf{0.8}  
} 
\vspace{\hdls} \\
\arrayrulecolor{ieee-dark-black-40}\hdashline\noalign{\vskip \hdls}
\multirow{3}{*}{\rot{\makecell[cc]{\vspace{-0.5em} \\ Circ. \\[-0.25em] Size \\}}} &
Transistors &
0.5B &
\dk &
\bs{39B} &
\bs{78B}\tn{k} &
\vf{8.8B} &
\vf{54B} &
\vf{80B} &
\dk &
\bs{4.8B} \\
&
Die Area [\sqmm] &
24.0&
585\tn{i} &
\bs{592}\tn{k} &
\vf{840}\tn{k} &
\vf{420} &
\vf{826} &
\vf{814} &
1600 &
\vf{146} \\
&
\scalebox{.91}[1.0]{Norm. Comp. Area [\sqmm]\tn{l}} &
3.86 (0.48/Core) &
292 (4.6/Core) &
\sh{442} (\sh{6.9}/Core) &
\sh{1183} (\sh{12.3}/Core) &
\sh{243} (\sh{5.1}/Core) &
\sh{632} (\sh{5.9}/SM) &
\sh{1170} (\sh{8.9}/SM) &
\sh{1024} (\sh{17}/Core) &
\sh{51.5} (\vf{1.0}/Cluster) 
\vspace{\hdls} \\
\arrayrulecolor{ieee-dark-black-40}\hdashline\noalign{\vskip \hdls}
\multirow{2}{*}{\rot{\makecell[cc]{\\ RAM \\}}} &
System DRAM & 
\dk &
$\leq$256GiB DDR4 &%
\vf{$\leq$4TiB} DDR4 &
\vf{$\leq$6TiB} DDR5 &
\vf{32GiB} HBM2 & 
\vf{40GiB} HBM2 & 
\vf{80GiB} HBM3 & 
\makecell[cl]{$\leq$4TiB DDR5 \\ 64GiB HBM2E} &
\vf{32GiB} HBM2E \\ 
&
On-Chip SRAM &
4.5 MiB &
64MiB L3\$ &
\vf{128MiB} L3\$ &
\vf{384MiB} L3\$ &
\vf{32MiB} L2\$ &
\vf{40MiB} L2\$ &
\vf{50MiB} L2\$ &
$\leq$112.5MiB L3\$ &
\bs{9MiB} SPM 
\vspace{\hdls} \\
\arrayrulecolor{ieee-dark-black-40}\hdashline\noalign{\vskip \hdls}
\multirow{1}{*}{\rot{\makecell[cc]{\vspace{-0.5em} \\ D2D \\[-0.25em] Link}}} &
\makecell[cl]{%
Technology \\
\#Links \\
\#Wires/Link \\
R+W BW $[$GB/s$]$ \\
\scalebox{.90}[1.0]{R+W BW/Link $[$GB/s$]$} \\
E. Eff $[$pJ/bit$]$
} &
\dk &
\dk &
\makecell[cl]{%
IFOP\tn{m}\;~\cite{amd_chiplet_isscc20} \\
8 \\
72 \\
440 \\
55/Link \\
2.0
} &
\makecell[cl]{%
IFOP\tn{m}\;~\cite{ravi2024zen4micro} \\
12 \\
\dk \\
1044 \\
87/Link \\
\dk
} &
\dk &
\dk &
\dk &
\makecell[cl]{%
EMIB\tn{b}\;\,/MDFIO\tn{n}~\cite{sapphire_rapids_isscc22} \\
20 \\
\dk \\
\textbf{10000} \\
\textbf{500/Link} \\
\textbf{0.5}
} &
\makecell[cl]{%
Fully digital DDR \\
38+1 \\
18 \\
16\tn{o} \\
0.42/Link \\
1.6\tn{o}
} 
\vspace{\hdls} \\
\arrayrulecolor{ieee-dark-black-40}\hdashline\noalign{\vskip \hdls}
\multirow{1}{*}[0.39cm]{\rot{\makecell[cc]{Peak \\[-0.25em] Compute \\[-0.25em] Throughp.}}} &
\makecell[cl]{%
FP Formats \\
$[$TFLOP/s$]$ \\
$[$GFLOP/s/\sqmm$]$ \\
$[$GFLOP/s/\sqmm$]$\tn{l}
} &
\makecell[cl]{%
\emph{FP64/32/16} \\
0.092/0.18/0.37 \\
20.9/41.8/83.7 \\
23.9/47.7/95.4
} &
\makecell[cl]{%
\emph{FP64/32/16} \\
1.00/2.00/4.00\hspace{0.1em}\cite{chen2020xuantie} \\
3.42/6.84/13.7 \\
3.42/6.84/13.7
} &
\makecell[cl]{%
\emph{FP64/32} \\
\sh{2.30\tn{p}\;\,/4.61\tn{p}} \\
\sh{8.22\tn{p}\;\,/16.4\tn{p}} \\
\sh{5.21\tn{p}\;\,/10.4\tn{p}}
} &
\makecell[cl]{%
\emph{FP64/32} \\
\sh{5.53\tn{p}\;\,/11.1\tn{p}} \\
\sh{11.8\tn{p}\;\,/23.6\tn{p}} \\
\sh{4.67\tn{p}\;\,/9.35\tn{p}}
} &
\makecell[cl]{%
\emph{FP64/32/16} \\
\vf{3.38/6.76/13.5} \\
\sh{21.9/43.9/87.8} \\
\sh{13.9/27.9/55.7}
} &
\makecell[cl]{%
\emph{FP64/32/16} \\
\vf{19.5\tn{q}\;\,/156\tn{q}\;\,/312\tn{q}} \\
\sh{48.6\tn{q}\;\,/389\tn{q}\;\,/777\tn{q}} \\
\sh{30.8\tn{q}\;\,/247\tn{q}\;\,/493\tn{q}}
} &
\makecell[cl]{%
\emph{FP64/32/16/8} \\
\textbf{66.9}\tn{q}\;\,/\textbf{495}\tn{q}\;\,/\textbf{989}\tn{q}\;\,/\textbf{1979}\tn{q}\;\, \\ %
\textbf{172}\tn{q}\;\,/\textbf{1270}\tn{q}\;\,/\textbf{2541}\tn{q}\;\,/\textbf{5082}\tn{q}\;\, \\
\textbf{57.2}\tn{q}\;\,/\textbf{423}\tn{q}\;\,/\textbf{845}\tn{q}\;\,/\textbf{1691}\tn{q}\;\,
} &
\makecell[cl]{%
\emph{FP64/32/16/8} \\
2.85\tn{r}~\cite{intel-app-metrics}/5.71/75.3/151\hspace{0.1em}\cite{intel-arch-day-2021} \\
4.84/9.67/128/255 \\
2.78/5.57/73.5/147
} &
\makecell[cl]{%
\emph{FP64/32/16/8} \\
\vf{0.77/1.54/3.07/6.14} \\
\sh{14.9/29.8/59.6/119} \\
\sh{14.9/29.8/59.6/119}
} 
\vspace{\hdls} \\
\arrayrulecolor{ieee-dark-black-40}\hdashline\noalign{\vskip \hdls}
\multirow{1}{*}[0.35cm]{\rot{\makecell[cc]{Best \\[-0.25em] Dense \\[-0.25em] LA FP64}}} &
\makecell[cl]{%
$[$GFLOP/s$]$ \\
FPU util. \\
$[$GFLOP/s/W$]$ \\
$[$GFLOP/s/\sqmm$]$ \\
$[$GFLOP/s/\sqmm$]$\tn{l}
} &
\makecell[cl]{%
87.5\tn{s}~\cite{schmidt2022rsoc, schmidt2018hwacha} \\
\textbf{\SI{95}{\percent}}\tn{t}~\cite{schmidt2018hwacha} \\
20.0\tn{s}~\cite{schmidt2022rsoc} \\
19.9\tn{s}~\;\cite{schmidt2022rsoc} \\
22.7\tn{s}~\;\cite{schmidt2022rsoc}
} &
\makecell[cl]{%
\emph{\dk} \\
\dk \\
\dk \\
\dk
} &
\makecell[cl]{%
1600\tn{p}\,~\cite{a64fx_dense} \\
\SI{70}{\percent}\tn{p}\,~\cite{a64fx_dense} \\
\dk \\
5.71\tn{p}\;\cite{a64fx_dense} \\
3.62\tn{p}\;\cite{a64fx_dense}
} &
\makecell[cl]{%
4428~\cite{amd2024tune9004} \\
\SI{80}{\percent}~\cite{amd2024tune9004} \\
\dk \\
9.43~\cite{amd2024tune9004} \\
3.74~\cite{amd2024tune9004}
} &
\makecell[cl]{%
1978\tn{u}\,~\cite{siegmann2024first} \\
\SI{72}{\percent}\tn{u}\,~\cite{siegmann2024first} \\
16.9\tn{u}\,~\cite{a64fx_isscc22} \\
12.8\tn{u}\;\cite{siegmann2024first} \\
8.15\tn{u}\;\cite{siegmann2024first}
} &
\makecell[cl]{%
18500~\cite{nvidia_a100_dense} \\
\SI{95}{\percent}~\cite{nvidia_a100_dense} \\
41.4\tn{x,y}\;\;\,\cite{strohmaier2024green500} \\
46.1~\cite{nvidia_a100_dense} \\
\textbf{29.3}~\cite{nvidia_a100_dense}
} &
\makecell[cl]{%
\textbf{33348}\tn{x,y,z}\;\;\;\;\,\cite{strohmaier2024top500}\cite{mellor2024virganews} \\
\SI{81}{\percent}\tn{x,y,z}\;\;\;\;\,\cite{strohmaier2024top500} \\
\textbf{65.4}\tn{x,y}\;\;\,\cite{strohmaier2024green500} \\
\textbf{85.7}\tn{x,y,z}\;\;\;\;\;\cite{strohmaier2024top500} \\
28.5\tn{x,y,z}\;\;\;\;\;\cite{strohmaier2024top500}
} &
\makecell[cl]{%
2696~\cite{siegmann2024first} \\
\SI{95}{\percent}~\cite{siegmann2024first} \\
\dk \\
4.57~\cite{siegmann2024first} \\
2.63~\cite{siegmann2024first}
} &
\makecell[cl]{%
686 \\
\SI{89}{\percent} \\
39.8 \\
13.3 \\
13.3
} 
\vspace{\hdls} \\
\arrayrulecolor{ieee-dark-black-40}\hdashline\noalign{\vskip \hdls}
\multirow{1}{*}[0.135cm]{\rot{\makecell[cc]{Best \\[-0.25em] Stencil \\[-0.25em] FP64}}} &
\makecell[cl]{%
$[$GFLOP/s$]$ \\
FPU util. \\
$[$GFLOP/s/W$]$ \\
$[$GFLOP/s/\sqmm$]$ \\
$[$GFLOP/s/\sqmm$]$\tn{l}
} &
\makecell[cl]{%
\dk \\
\dk \\
\dk \\
\dk \\
\dk
} &
\makecell[cl]{%
\emph{\dk} \\
\dk \\
\dk \\
\dk \\
\dk
} &
\makecell[cl]{%
855\tn{p}\,~\cite{cfd_amd_epyc_rome_tpds21} \\
\sh{\SI{37}{\percent}}\tn{p}\,~\cite{cfd_amd_epyc_rome_tpds21} \\
{\dk} \\
\sh{3.05}\tn{p}\;\cite{cfd_amd_epyc_rome_tpds21} \\
\sh{1.93}\tn{p}\;\cite{cfd_amd_epyc_rome_tpds21}
} &
\makecell[cl]{%
\dk \\
\dk \\
\dk \\
\dk \\
\dk
} &
\makecell[cl]{%
323\tn{u}\,~\cite{hirokawa2022large} \\
\sh{\SI{11}{\percent}}\tn{u}\,~\cite{hirokawa2022large} \\
{\dk} \\
\sh{2.10}\tn{u}\;\cite{hirokawa2022large} \\
\sh{1.33}\tn{u}\;\cite{hirokawa2022large}
} &
\makecell[cl]{
\rev{\textbf{6063}\,~\cite{chen2024convstencil}} \\
\sh{\SI{49}{\percent}}\tn{v}\,~\cite{nvidia_a100_stencil} \\
{\dk} \\
\rev{\sh{\bf{15.1}}\,\cite{chen2024convstencil}} \\
\rev{\sh{9.58}\,\cite{chen2024convstencil}}
} &
\makecell[cl]{%
\dk \\
\dk \\
\dk \\
\dk \\
\dk
} &
\makecell[cl]{%
\emph{\dk} \\
\dk \\
\dk \\
\dk \\
\dk
} &
\makecell[cl]{%
571 \\
\sh{\textbf{\SI{83}{\percent}}} \\
\sh{\textbf{28.1}} \\
\sh{11.1} \\
\sh{\textbf{11.1}}
} 
\vspace{\hdls} \\
\arrayrulecolor{ieee-dark-grey-60}\hdashline\noalign{\vskip \hdls}
\multirow{1}{*}[0.25cm]{\rot{\makecell[cc]{Best \\[-0.25em] Sp-dense \\[-0.25em] LA FP64}}} &
\makecell[cl]{%
$[$GFLOP/s$]$ \\
FPU util. \\
$[$GFLOP/s/W$]$ \\
$[$GFLOP/s/\sqmm$]$ \\
$[$GFLOP/s/\sqmm$]$\tn{l}
} &
\makecell[cl]{%
\dk \\
\dk \\
\dk \\
\dk \\
\dk
} &
\makecell[cl]{%
\emph{\dk} \\
\dk \\
\dk \\
\dk \\
\dk
} &
\makecell[cl]{%
\sh{187}\tn{p}\,~\cite{amd_sparse} \\
\sh{\SI{8.1}{\percent}}\tn{p}\,~\cite{amd_sparse} \\
\dk \\
\sh{0.67}\tn{p}\;\cite{amd_sparse} \\
\sh{0.42}\tn{p}\;\cite{amd_sparse}
} &
\makecell[cl]{%
67.5~\cite{amd2024tune9004} \\
\SI{1.2}{\percent}~\cite{amd2024tune9004} \\
\dk \\
0.14~\cite{amd2024tune9004} \\
0.06~\cite{amd2024tune9004}
} &
\makecell[cl]{%
\sh{131}\tn{u}\,~\cite{alappat2020a64fx} \\
\sh{\SI{4.7}{\percent}}\tn{u}\,~\cite{alappat2020a64fx} \\
\dk \\
\sh{0.85}\tn{u}\;\cite{alappat2020a64fx} \\
\sh{0.54}\tn{u}\;\cite{alappat2020a64fx}
} &
\makecell[cl]{%
\sh{286}\tn{v}\,~\cite{nvidia_a100_sparse} \\
\sh{\SI{2.9}{\percent}}\tn{v}\,~\cite{nvidia_a100_sparse} \\
\dk \\
\sh{0.71}\tn{v}\;\cite{nvidia_a100_sparse} \\
\sh{0.45}\tn{v}\;\cite{nvidia_a100_sparse}
} &

\makecell[cl]{%
\emph{\dk} \\
\dk \\
\dk \\
\dk \\
\dk
} &
\makecell[cl]{%
\sh{198}~\cite{siegmann2024first} \\
\SI{6.9}{\percent}~\cite{siegmann2024first} \\
{\dk} \\ 
0.33~\cite{siegmann2024first} \\
0.19~\cite{siegmann2024first}
} &
\makecell[cl]{%
\sh{\textbf{307}} \\
\sh{\textbf{\SI{42}{\percent}}} \\
\sh{\textbf{16.0}} \\
\sh{\textbf{5.95}} \\
\sh{\textbf{5.95}}
} 
\vspace{\hdls}\\
\arrayrulecolor{black}\bottomrule
\end{tabular}
\begin{tablenotes}[para, flushleft]
    \im{a} Chip on Wafer on Substrate
    \im{b} Embedded Multi-Die Interconnect Bridges
    \im{c} Based on core presented in~\cite{chen2020xuantie}
    \im{d} Interposer: 65nm
    \im{e} 512 bit
    \im{f} Xoc := Xidma\_Xsssr\_Xminifloat
    \im{g} Host: RV64GC
    \im{h} RV64GC part of core as \emph{OpenC910}
    \im{i} Measured on released product
    \im{k} w/o IO die
    \im{l} Scaled to 12LP+ node using NAND2X1 gate areas (\si{GE\per\micro\metre\squared}) or transistor density
    \im{m} Infinity-Fabric On-Package
    \im{n} Multi-Die Fabric I/O
    \im{o} Wide segment (38 links) 
    \im{p} At base (non-boost) clock
    \im{q} Fully utilized tensor cores
    \im{r} At 2.6 \si{\giga\hertz}
    \im{s} At nominal 1V supply
    \im{t} Based on vector unit only
    \im{u} At below-peak clock \rev{(FPU util. relative to peak at reduced clock)}
    \im{v} Tensor cores unused (FPU util. relative to peak without tensor cores)
    \im{x} Unknown whether tensor cores used
    \im{y} Based on many-GPU supercomputers
    \im{z} Based on "Virga" system using 448 H100 GPUs~\cite{mellor2024virganews}
\end{tablenotes}
\end{threeparttable}
} %
\end{table}
\end{landscape}%
}

On sparse-dense FP64 \gls{la}, Occamy's indirect \glspl{su} and explicitly managed memory hierarchy enable even higher utilization gains.
Although all competitors have \gls{isa}-level support for scatter-gather, they reach at most \cn{\SI{8.1}{\percent}} FPU utilization on AMD Rome~\cite{amd_sparse}, \cn{5.2\x}~less than the \cn{\SI{42}{\percent}} we achieve;
this highlights the importance of our explicitly managed dataflow.
We also achieve the highest technology-node-normalized compute density of \cn{\SI{5.95}{DP\text{-}\giga\flop\per\second\per\milli\meter^2}}, \cn{11\x}~higher than the leading competitor A64FX~\cite{alappat2020a64fx}. %
While A100 introduces hardware support for $2$:$4$ structured sparsity ($\leq$2 nonzeros each block of four elements)~\cite{nvidia2020a100tc}, this feature cannot be used in~\cite{nvidia_a100_sparse} as it is specifically designed for deep learning on $\leq$\SI{32}{\bit} datatypes and ineffective in processing the highly sparse ($\gtrsim$\SI{95}{\percent}) data found in graph \gls{ml} and \gls{hpc} applications.
As a result, A100's inflexible GPU datapath achieves at most \cn{\SI{2.9}{\percent}} \gls{fpu} utilization, which is less than more agile CPU architectures we compare to.

To summarize, Occamy is the first 2.5D-integrated multi-chiplet \riscv~manycore demonstrated in silicon.
Compared to state-of-the-art CPUs and GPUs, it achieves competitive FPU utilization (\cn{\SI{89}{\percent}}) on dense \gls{la}, as well as leading FPU utilizations (\cn{1.7\x}, \cn{5.2\x}) and technology-node-normalized compute densities (\rev{\cn{1.2\x}}, \cn{11\x}) on stencil codes and sparse-dense \gls{la}, respectively.
Unlike existing sparse compute accelerators and \gls{isa} extensions, our \glspl{su} are highly flexible and do not narrowly specialize to specific problem domains or data precisions.
As such, Occamy combines high dense efficiency, sparse efficiency, and architectural flexibility in an unprecedented way.

\section{Conclusion}
\label{sec:conlusion}

We presented {\occamy}, the first open-source silicon-proven 2.5D-integrated \riscv~manycore featuring 432 cores across two chiplets, \SI{32}{\gibi\byte} of HBM2E DRAM, and a fully-digital fault-tolerant \cn{\SI{1.6}{\pico\joule\per\bit}} \glsreset{d2d}\gls{d2d} link.
{\occamy}'s compute cores feature an 8-to-64-bit SIMD-capable FPU kept busy by two ISA extensions: a hardware loop and three sparsity-capable \glsreset{su}\glspl{su}.
Our \glspl{su} enable high FPU utilization on both sparse and dense workloads through $\leq$4D affine and indirect streams as well as streaming intersection and union.
Agile, latency-tolerant DMA engines copy large data tiles to and from local \glsreset{spm}\glspl{spm} to be processed by the compute cores with \glspl{su}.

We described the hierarchical implementation of our \SI{12}{\nano\meter} FinFET compute chiplets and our \SI{65}{\nano\meter} interposer \emph{Hedwig}.
On dense \glsreset{la}\gls{la}, {\occamy} achieves a competitive FPU utilization of up to \cn{\SI{89}{\percent}} and an energy efficiency of up to \cn{\SI{39.8}{DP\text{-}\giga\flop\per\second\per\watt}}.
On stencil codes, we reach an FPU utilization of up to \cn{\SI{83}{\percent}} and a technology-node-normalized compute density of up to \cn{\SI{11.1}{DP\text{-}\giga\flop\per\second\per\milli\meter^2}}, improving by \cn{1.7\x}~and \rev{\cn{1.2\x}}, respectively, over \gls{soa} CPUs and GPUs.
On sparse-dense \gls{la}, we achieve up to \cn{\SI{42}{\percent}} FPU utilization and a technology-node-normalized compute density of up to \cn{\SI{5.95}{DP\text{-}\giga\flop\per\second\per\milli\meter^2}}, resulting in substantial gains of \cn{5.2\x}~and \cn{11\x} over the \gls{soa}.
We \rev{further} achieve a sparse-sparse \gls{la} throughput of up to \cn{\SI{187}{\giga\comp\per\second}} at \cn{\SI{17.4}{\giga\comp\per\second\per\watt}}, \cn{\SI{3.63}{\giga\comp\per\second\per\milli\meter^2}}, and a comparator utilization of \cn{\SI{49}{\percent}}.
\rev{Finally, we reach FPU utilizations of up to \SI{75}{\percent} and \SI{54}{\percent} on 
\gls{gpt}-J inference and a \gls{gcn} layer, respectively.} 
The \gls{rtl} description of {\occamy}'s entire compute architecture is publicly available under a permissive open-source license\footnote{\url{https://github.com/pulp-platform/occamy}}.

\section*{Acknowledgments}
\rev{We thank the reviewers for their valuable feedback.}
\noindent
We thank 
Florian Zaruba, Fabian Schuiki, Samuel Riedel, Alfio Di Mauro, Stefan Mach, Sina Arjmandpour, Noah H\"utter, Andreas Kurth, 
Alfonso Fontao, Beat Muheim, Zerun Jiang, 
Georg Rutishauser, Silvio Scherr, Aldo Rossi,
Cyril Koenig,
Michael Rogenmoser, Cristian Cioflan, and Philippe Sauter
for their invaluable contributions to this project.
We are deeply grateful to GlobalFoundries, Avery, Micron, Rambus, and Synopsys for their generous support.

This work was supported in part through funding from the European High Performance Computing Joint Undertaking (JU) under Framework Partnership Agreement No 800928 and
Specific Grant Agreement No: 101036168 (EPI SGA2) and No: 101034126 (The EU Pilot).
%

%

%

%

%
\newcommand{\missingbio}{has not yet added their bio. This is just a placeholder.}
\newcommand{\lucaphd}[1]{#1 is currently pursuing a Ph.D. degree in the Digital Circuits and Systems group of Prof.\ Benini.}
\newcommand{\ethgrad}[3]{received #1 B.Sc. and M.Sc. degrees in electrical engineering and information technology from ETH Zurich in #2 and #3, respectively.}
\newcommand{\lucagrad}[2]{completed #1 Ph.D. in the Digital Circuits and Systems group of Prof.\ Benini in #2.}
\newcommand{\researchinterests}[1]{research interests include #1.}

\begin{IEEEbiography}[%
    {\includegraphics[width=1in,height=1.25in,clip,keepaspectratio]{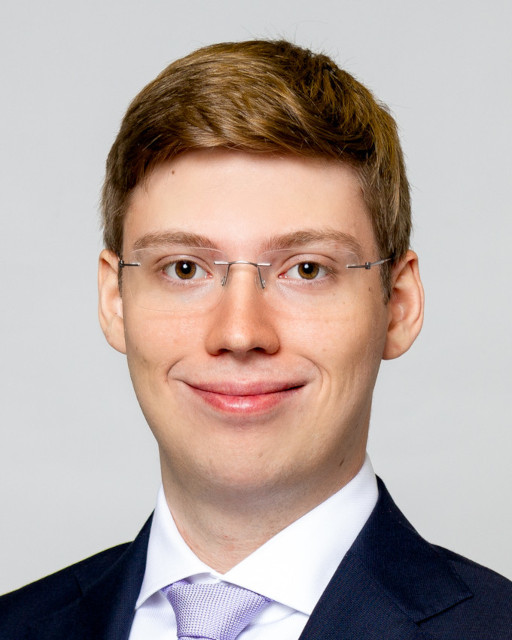}}%
    ]{Paul Scheffler}
    (Graduate Student Member, IEEE) 
    \ethgrad{his}{2018}{2020}
    \lucaphd{He}
    His
    \researchinterests{hardware acceleration of sparse and irregular workloads, on-chip interconnects, manycore architectures, and high-performance computing}
\end{IEEEbiography}

\begin{IEEEbiography}[%
    {\includegraphics[width=1in,height=1.25in,clip,keepaspectratio]{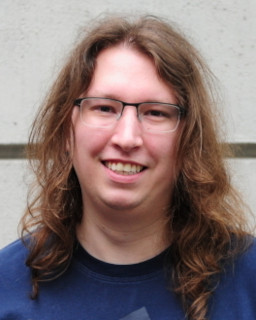}}%
    ]{Thomas Benz}
    (Graduate Student Member, IEEE) 
    \ethgrad{his}{2018}{2020}
    \lucaphd{He}
    His
    \researchinterests{energy-efficient high-performance computer architectures, memory interconnects, data movement, and the design of \acrshortpl{asic}}
\end{IEEEbiography}

\begin{IEEEbiography}[%
    {\includegraphics[width=1in,height=1.25in,clip,keepaspectratio]{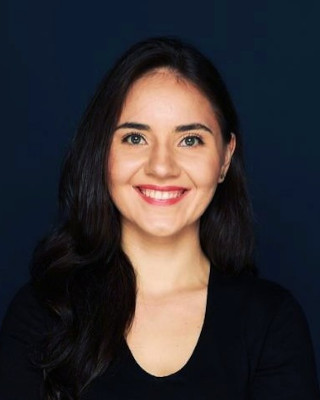}}%
    ]{Viviane Potocnik}
    (Graduate Student Member, IEEE)
    \ethgrad{her}{2020}{2022}
    \lucaphd{She}
    Her
    \researchinterests{heterogeneous architectures and the exploration of innovative data representation strategies to enhance the computational efficiency and adaptability on devices at the extreme edge}
\end{IEEEbiography}

\begin{IEEEbiography}[%
    {\includegraphics[width=1in,height=1.25in,clip,keepaspectratio]{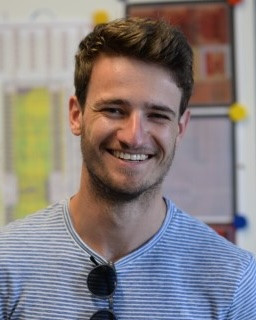}}%
    ]{Tim Fischer}
    (Graduate Student Member, IEEE) 
    \ethgrad{his}{2018}{2021}
    \lucaphd{He}
    His
    \researchinterests{scalable and energy-efficient on-chip and off-chip interconnects for high-performance computing}
\end{IEEEbiography}

\begin{IEEEbiography}[%
    {\includegraphics[width=1in,height=1.25in,clip,keepaspectratio]{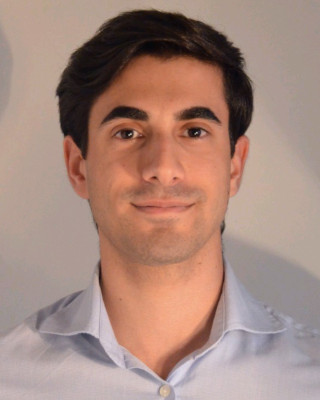}}%
    ]{Luca Colagrande}
    (Graduate Student Member, IEEE) 
    received his B.Sc. degree from Politecnico di Milano in 2018 and his M.Sc. degree from ETH Zurich in 2020.
    \lucaphd{He}
    His
    \researchinterests{energy-efficient general-purpose manycore accelerators and hardware-software co-design for machine learning and high-performance computing applications}
\end{IEEEbiography}

\begin{IEEEbiography}[%
    {\includegraphics[width=1in,height=1.25in,clip,keepaspectratio]{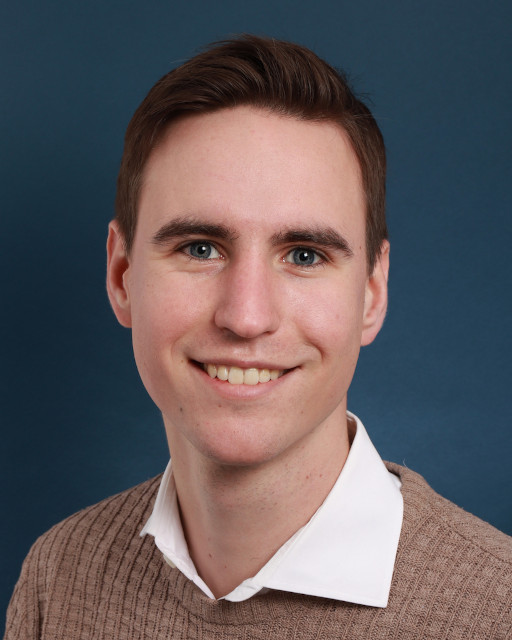}}%
    ]{Nils Wistoff}
    (Graduate Student Member, IEEE)
    received his B.Sc. and M.Sc. degrees from RWTH Aachen University in 2017 and 2020, respectively. 
    \lucaphd{He}
    His
    \researchinterests{processor and system-on-chip design and secure computer architecture}
\end{IEEEbiography}

\begin{IEEEbiography}[%
    {\includegraphics[width=1in,height=1.25in,clip,keepaspectratio]{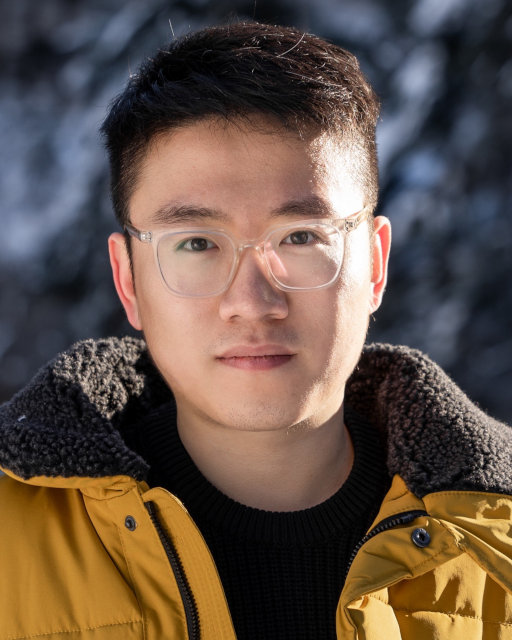}}%
    ]{Yichao Zhang}
    (Graduate Student Member, IEEE)
    received his B.Eng. degree from Heilongjiang University China in 2015 and his M.Sc. degree from Nanyang Technological University Singapore in 2017.
    \lucaphd{He}
    His
    \researchinterests{physically feasible manycore RISC-V architectures, parallel computing, and SIMD processing}
\end{IEEEbiography}

\begin{IEEEbiography}[%
    {\includegraphics[width=1in,height=1.25in,clip,keepaspectratio]{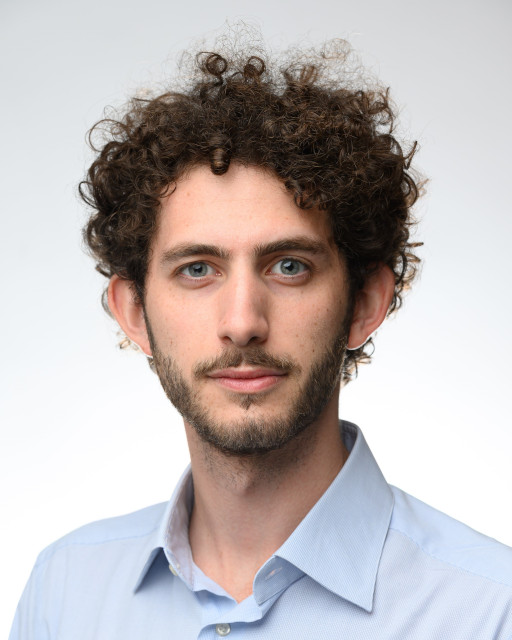}}%
    ]{Luca Bertaccini}
    (Graduate Student Member, IEEE)
    received his M.Sc. degree in Electronic Engineering from the University of Bologna in 2020. 
    \lucaphd{He}
    His
    \researchinterests{heterogeneous systems-on-chip, energy-efficient hardware accelerators, computer arithmetic, and mixed-precision computing}
\end{IEEEbiography}

\begin{IEEEbiography}[%
    {\includegraphics[width=1in,height=1.25in,clip,keepaspectratio]{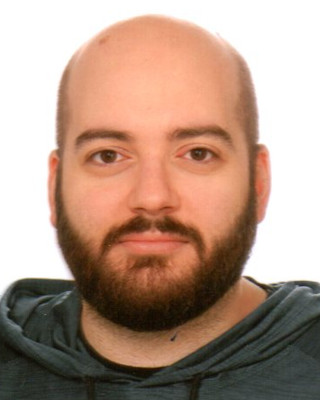}}%
    ]{Gianmarco Ottavi}
    received his M.Sc. degree from the University of Bologna in 2019. He has worked as a research fellow and is currently pursuing his Ph.D. at the Department of Electrical, Electronic, and Information Engineering "Guglielmo Marconi" at the University of Bologna. His research interests include energy-efficient architectures for DNN inference, vector machines, and computer architectures.
\end{IEEEbiography}

\begin{IEEEbiography}[%
    {\includegraphics[width=1in,height=1.25in,clip,keepaspectratio]{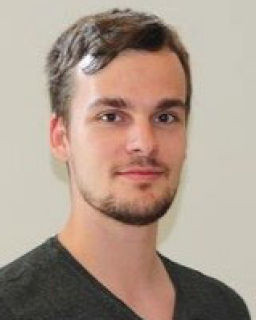}}%
    ]{Manuel Eggimann}
    (Member, IEEE) 
    \ethgrad{his}{2016}{2018}
    He
    \lucagrad{his}{2023}
    His
    \researchinterests{low-power hardware design, edge computing, and VLSI}
\end{IEEEbiography}

\begin{IEEEbiography}[%
    {\includegraphics[width=1in,height=1.25in,clip,keepaspectratio]{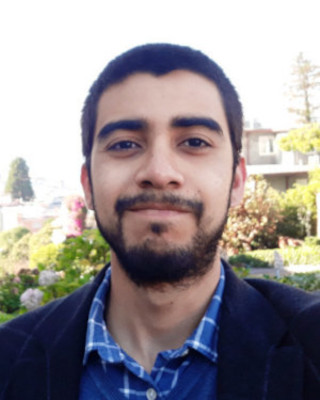}}%
    ]{Matheus Cavalcante}
    (Member, IEEE)
    received his M.Sc. degree in Integrated Electronic Systems from the Grenoble Institute of Technology (Phelma), France, in 2018. 
    He
    \lucagrad{his}{2023}
    His
    \researchinterests{vector processing, high-performance computer architectures, and emerging VLSI technologies}
\end{IEEEbiography}

\begin{IEEEbiography}[%
    {\includegraphics[width=1in,height=1.25in,clip,keepaspectratio]{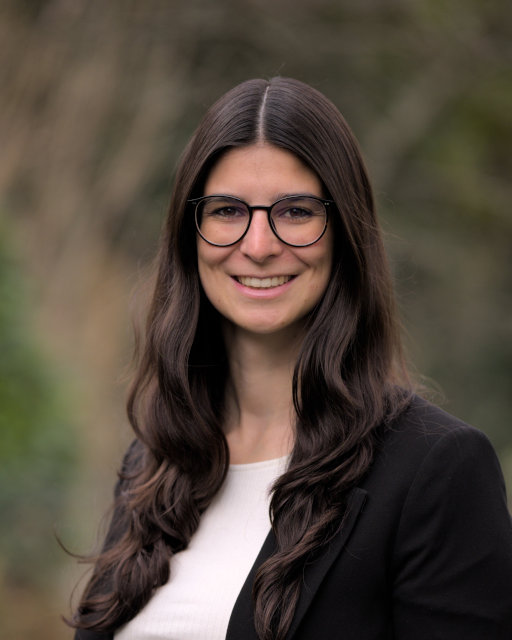}}%
    ]{Gianna Paulin}
    (Member, IEEE)    
    \ethgrad{her}{2018}{2021}
    She
    \lucagrad{her}{2023}
    She worked on energy-efficient compute architectures accelerating neural networks and was involved in more than seven tape-outs. %
    In March 2024, she joined Axelera AI, where she is developing the next generation of AI accelerators. 
\end{IEEEbiography}

\vfill
\newpage

\begin{IEEEbiography}[%
    {\includegraphics[width=1in,height=1.25in,clip,keepaspectratio]{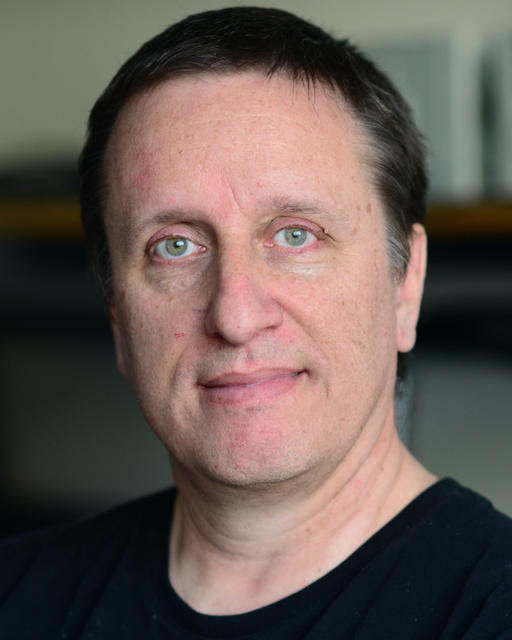}}%
    ]{Frank K. G\"{u}rkaynak}
    (Member, IEEE) has obtained his B.Sc. and M.Sc. in electrical engineering from the Istanbul Technical University, and his Ph.D. in electrical engineering from ETH Z\"urich in 2006. He is currently working as a senior scientist at the Integrated Systems Laboratory of ETH Z\"urich. His research interests include digital low-power design and cryptographic hardware.
\end{IEEEbiography}

\begin{IEEEbiography}[%
    {\includegraphics[width=1in,height=1.25in,clip,keepaspectratio]{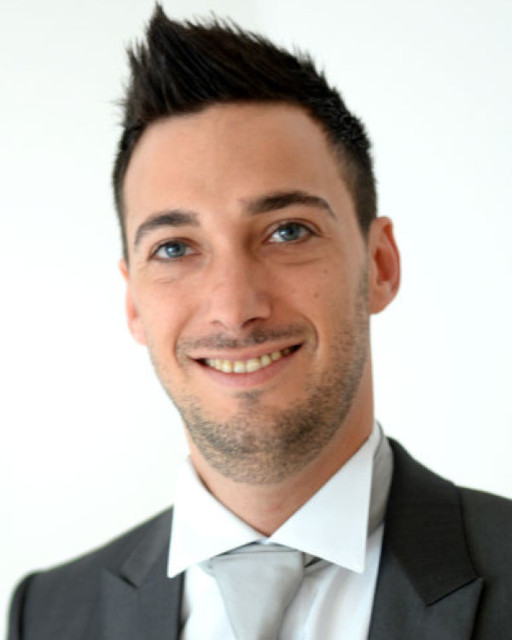}}%
    ]{Davide Rossi}
    (Senior Member, IEEE) received the Ph.D. degree from the University of Bologna, Bologna, Italy, in 2012. He has been a Post-Doctoral Researcher with the Department of Electrical, Electronic and Information Engineering “Guglielmo Marconi,” University of Bologna, since 2015, where he is currently an Associate Professor. His research interests include energy-efficient digital architectures. 
\end{IEEEbiography}

\begin{IEEEbiography}[%
    {\includegraphics[width=1in,height=1.25in,clip,keepaspectratio]{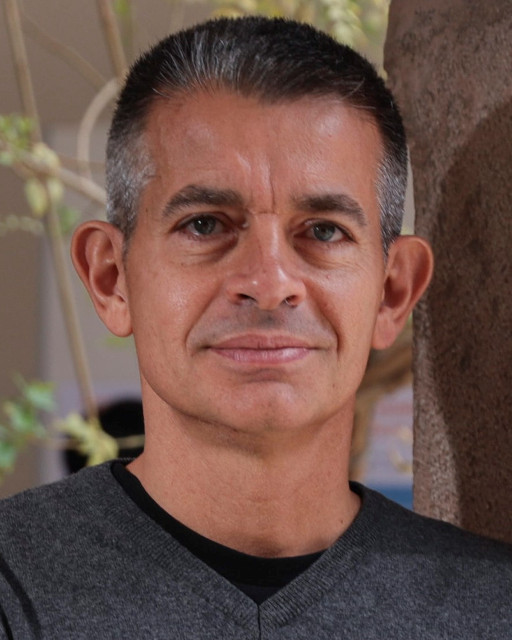}}%
    ]{Luca Benini}
    (Fellow, IEEE) holds the chair of Digital Circuits and Systems at ETH Zurich and is Full Professor at the Università di Bologna.
    Dr.\ Benini’s research interests are in energy-efficient computing systems design, from embedded to high-performance.
    He has published more than 1000 peer-reviewed papers and five books.
    He is a Fellow of the ACM and a member of Academia Europaea.
\end{IEEEbiography}

\vfill

\end{document}